\documentclass[referee,a4paper,12pt,traditabstract]{jswsc} 

\usepackage{mathtools} 
\usepackage{graphicx}
\usepackage{txfonts}
\usepackage{subfigure}
\usepackage{epstopdf}
\usepackage[mathlines]{lineno} 
\usepackage[authoryear,round]{natbib}
\usepackage[backref]{hyperref}
\usepackage{url}
\usepackage{float} 

\hypersetup{colorlinks=true,citecolor=cyan,urlcolor=cyan,linkcolor=blue}

\begin{document}

   \title{Nowcasting Geoelectric Fields in Ireland using Magnetotelluric Transfer Functions}
   
   \titlerunning{Nowcasting Geoelectric Fields using MT-TF}

   \authorrunning{Malone-Leigh et al}

\author{John Malone-Leigh* (\href{mailto:jmalone@cp.dias.ie}{jmalone@cp.dias.ie})
        \inst{1,2}
        \and
        Joan Campany\`{a}
        \inst{3,4}
        \and
        Peter T. Gallagher
        \inst{1}
        \and
        Maik Neukirch
        \inst{5}
        \and
        Colin Hogg
        \inst{6}
        \and
        Jim Hodgson
        \inst{4}
        }

\institute{Astronomy \& Astrophysics Section, DIAS Dunsink Observatory, Dublin Institute for Advanced Studies, Dublin, Ireland
            \and
            School of Physics, Trinity College Dublin, Dublin, Ireland
            \and
            South East Technological University, Carlow,  Ireland
            \and
            Geological Survey Ireland, Dublin, Ireland
            \and
            Department of Geophysics, University of Oslo, Oslo, Norway
            \and
            Geophysics Section, Dublin Institute for Advanced Studies, Dublin, Ireland
             }
  \date{Received 02 Aug 2022; accepted 07 Feb 2023}
 
  \abstract
   {Geomagnetically induced currents (GIC) driven by geoelectric fields pose a hazard to ground-based infrastructure, such as power grids and pipelines. Here, a new method is presented for modelling geoelectric fields in near real time, with the aim of providing valuable information to help mitigate the impact of GIC. The method uses magnetic field measurements from the Magnetometer Network of Ireland (MagIE; \url{www.magie.ie}), interpolates the geomagnetic field variations between magnetometers using spherical elementary current systems (SECS), and estimates the local electric field using a high density ($<~40~km$) network of magnetotelluric transfer functions (MT-TF) encompassing the island. The model was optimised to work in near real time, with a correction curve applied to the geoelectric field time series. This approach was successfully validated with measured electric fields at four sites for a number of geomagnetic storms, providing accurate electric fields up to a 1-minute delay from real time, with high coherence (0.70 -- 0.85) and signal-to-noise ratio (SNR; 3.2 -- 6.5) relative to measured electric field validation time series .This was comparable to a standard non real-time geoelectric field model (coherence~=~0.80 -- 0.89 and SNR~=~4.0 -- 7.0). The impact of galvanic distortion on the model was also briefly evaluated, with a galvanic distortion correction leading to a more homogeneous representation of the direction of the electric field, at a regional scale. }


   \maketitle


\section{Introduction}\label{section:Introduction}
    Magnetic variations originating from currents in the magnetosphere and ionosphere are responsible for induced geoelectric fields at the Earth’s surface. Geoelectric fields are a significant hazard to ground-based technology, being responsible for geomagnetically induced currents (GIC) in pipelines and power grids. Many examples exist of damage to transformers and disruption to power networks, such as the well known example of Quebec, Canada in 1989, as well other examples such as New Zealand in November, 2001 \citep{{Allen1989}, {Beland2004}, {Bolduc2002}} and South Africa and Sweden in October, 2003 \citep{{Gaunt2005}, {Koen2003}, {Pulkkinen2005}}. In Ireland, \citet{Blake2016} and \citet{Blake2018a} evaluated the consequences of severe geomagnetic storms on the Irish power grid. They predicted a potentially large GIC in the south-west of the island. Accurate real-time assessment and forecasting of geoelectric fields could aid in reducing the hazard posed by some of these storms.

    \paragraph{}
    To monitor the potential impacts caused by geomagnetic storms on ground-based infrastructure, knowledge about the intensity and direction of the geoelectric fields is required. This can be achieved by either measuring or modelling the geoelectric fields. Modelling geoelectric fields is the practical approach for two main reasons: 1) Geological variations resulting in substantial differences in geoelectrical properties at lithospheric depth (the outermost shell of the Earth made up of the crust and a portion of upper mantle) and can lead to significant differences in the geoelectric fields between nearby stations. The complex distribution of the geoelectrical properties of Ireland’s lithosphere requires many measuring electrometers, a network which would be challenging and expensive to operate. 2) Real-time measurements of geoelectric fields are highly sensitive to localised noise from electrical and ferrous sources \citep{Schmidt2020}. While modelling is preferred for real-time monitoring of geoelectric fields, measurements of geoelectric fields are key to validating the model. 
    
    \paragraph{}

    Magnetotelluric transfer functions (MT-TF) are an effective approach for modelling geoelectric fields, suitable to the complexity of the geoelectrical structures beneath the island of Ireland \citep{Rao2014}. MT-TF relate magnetic variations to electric field variations using MT-TF at individual sites, related to the subsurface resistivity of the lithosphere. The use of 3D resistivity models and 3D MT-TF produced highly accurate results, when compared to other common geoelectric field models, such as simpler 1D resistivity profiles and the thin sheet model \citep{{Weigel2017}, {Beggan2021}}. MT-TF were chosen for this study over traditional 3D lithospheric models, because of their computational efficiency, as they only use a transfer function and do not require a full resistivity model of the lithosphere. This is important when nowcasting, as a time efficient method is required to ensure the model is truly real time. However, recent research by \citet{Kruglyakov2022} presents a possible alternative to this problem, by using a memory-based method to significantly reduce computation time of 3D resistivity models. 
    
    \paragraph{}
    
    Magnetic field data are monitored at specific geomagnetic observatories, and so, there is need for an interpolation approach to constrain the magnetic field variations at the MT sites used to model the geoelectric fields. A Spherical Elementary Current Systems (SECS) interpolation was developed to model magnetic field variations. The SECS method \citep{Amm1997,Amm+Viljanen} is commonly used to estimate magnetic field variations at sites without magnetic field data, and uses nearby magnetometers as inputs \citep{Marshalko,Beggan2021,Bosse} to construct a plane of equivalent sheet currents in the ionosphere, for a set altitude. More computationally efficient approaches were considered for estimating the geomagnetic field, such as magnetic perturbation functions \citep{SimpsonBahr2}. Given that Ireland is a small region, a simpler approach may produce accurate results and would speed up computation time. However, significant differences were observed between the north, south east and west of Ireland, in terms of amplitude and local effects during geomagnetic storms (\citet{Blake2018}, example in Fig. \ref{fig:SECS}) and as such, a more complex interpolation method was required to obtain reliable results. The SECS method works reliably for sites at higher latitudes, including similar latitudes to Ireland \citep{McLay}. It must be noted, however, that the accuracy of SECS interpolations fall off at middle and low latitudes \citep{Torta2017} and may require separate interpolation methods, or a modification of the SECS in these regions, to return accurate results \citep{Vanhamaki}.
    \paragraph{}
    
    Here, we combined the SECS-MT approach with the real-time FFT modelling method of \cite{Kelbert2017} and expand on it by applying an additional real-time correction. The nowcast geoelectric field was compared with and optimised using a standard non real-time model (Section \ref{section:correction_curve}), using MT-TF from Space Weather Electromagnetic Database for Ireland \citep[SWEMDI;][]{Campanya2018}. The optimisation involved obtaining a correction curve, which was used to correct the electric field time series near real time, to provide the most accurate electric fields at each instance, for generation of a real-time movie. The nowcasting approach, based on the SECS-MT modelling technique, was validated against measured geoelectric field data (Section \ref{section:validation}). In addition, the impact of galvanic distortions on modelling geoelectric fields was evaluated, comparing results with and without correcting the MT impedance tensors for galvanic distortion (\ref{section:Galvanic_dist}) following the approach of \citet{Neukirch2020}.

    \paragraph{}

    \section{Methods}
    \subsection{Geomagnetic Fields: SECS}\label{section:SECS}
\begin{figure}[H]
        \centering
        \includegraphics[scale=0.6]{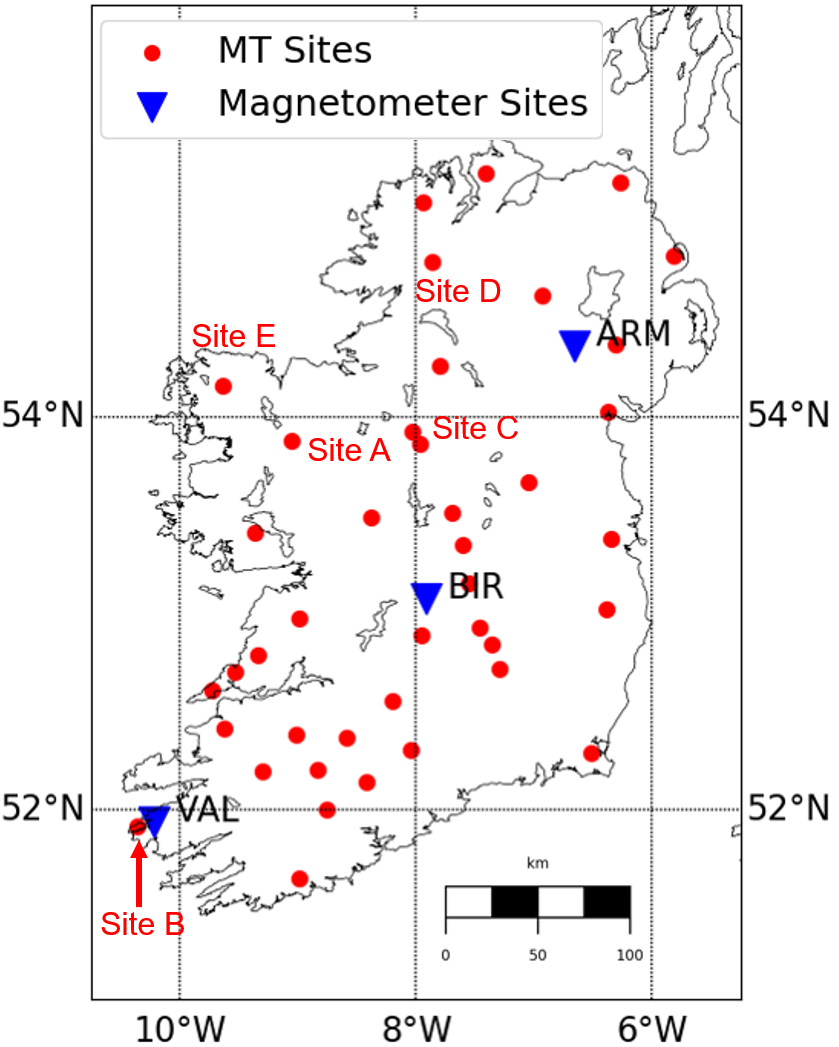}
        \caption{MagIE (ARM, BIR, VAL) magnetometer observatories (blue) and the 43 MT site locations (red) used in this study. Example MT sites (A, B, C, D, E) used subsequently, are marked with red text. Site A was an example site used when optimising a correction for the nowcast geoelectric field. Sites B - E were all used to validate the nowcast model.}
        \label{fig:Mag_MT_Sites}
    \end{figure}

\begin{figure}[H]
        \centering
        \includegraphics[scale=0.5]{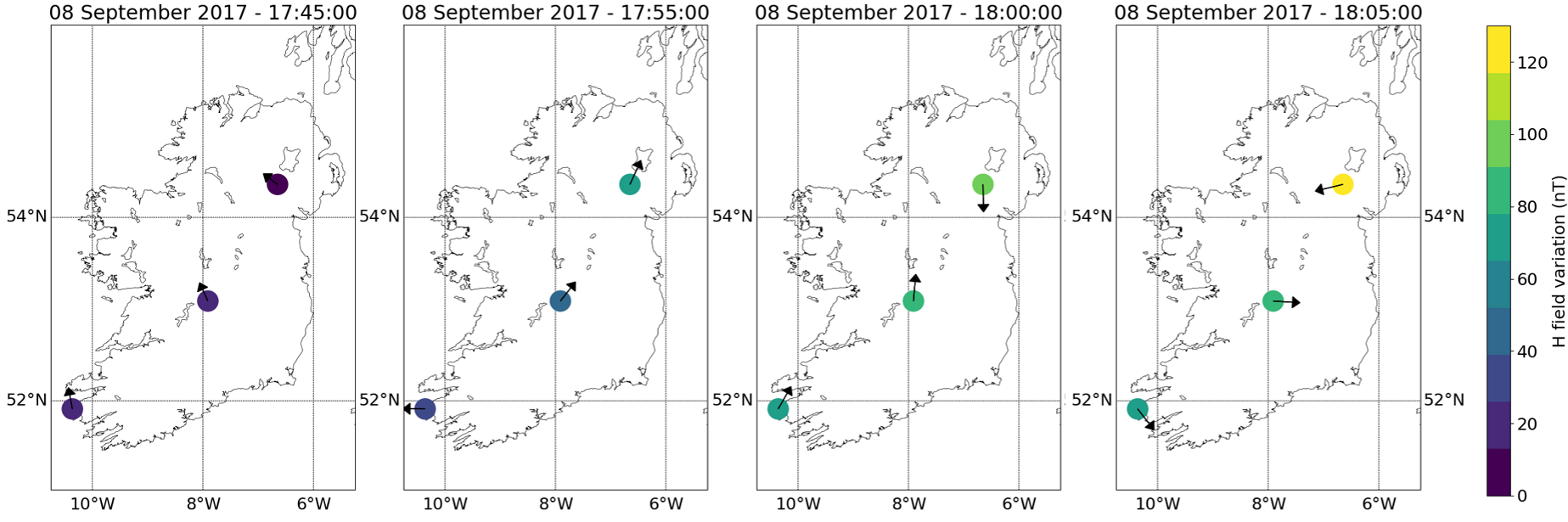}
        \includegraphics[scale=0.49]{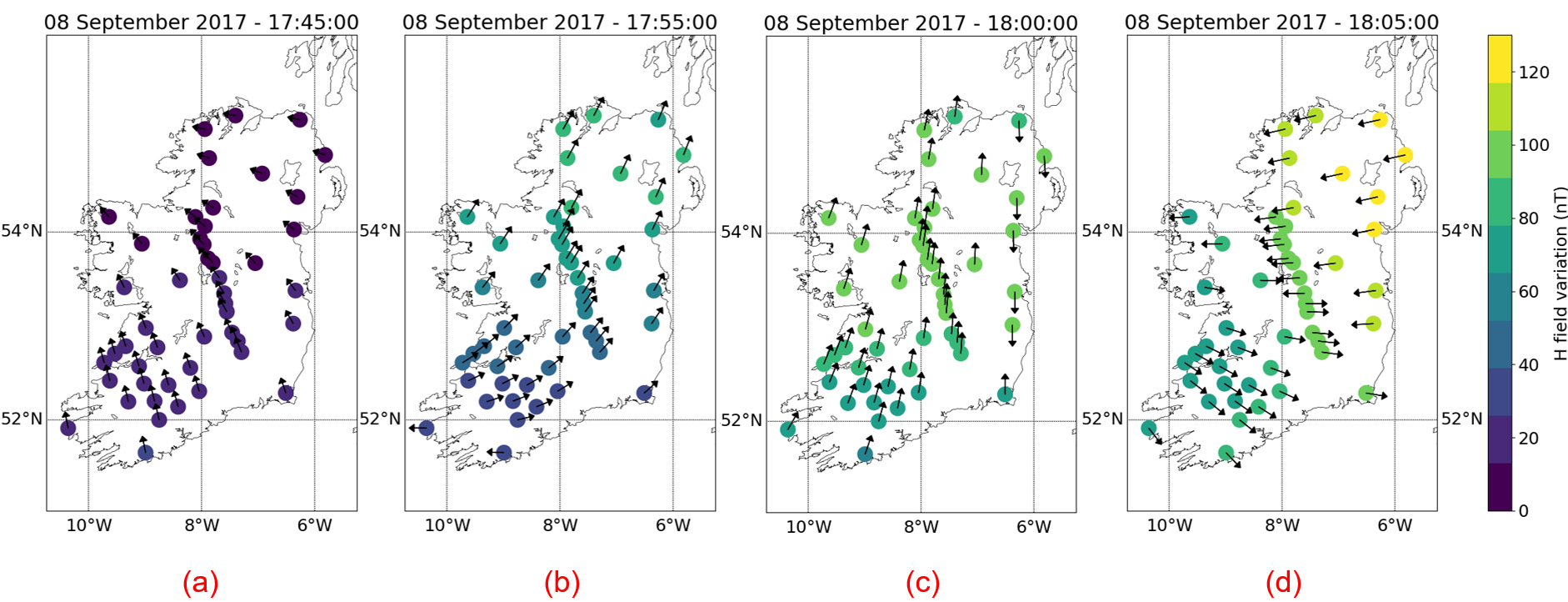}        
        \caption{(top) An example of the input magnetic fields at each magnetometer for the 08 September 2017, during an increase in geomagnetic activity. (bottom) The estimated magnetic field at MT sites using a SECS interpolation. At (a) no activity is present. Over time, activity increases from north-east to south-west, with the direction of magnetic fields changing.}
        \label{fig:SECS}
    \end{figure}

    Magnetic field variations are monitored in Ireland by the Magnetometer Network of Ireland (MagIE, \url{www.magie.ie}) at Armagh, Birr and Valentia observatories (Fig. \ref{fig:Mag_MT_Sites}). Magnetic time series (1~--~second resolution) from the magnetic observatories were used to estimate the magnetic field (Fig. \ref{fig:SECS}) at each of the MT sites in Figure \ref{fig:Mag_MT_Sites} using a SECS interpolation. The SECS interpolation was implemented across Ireland using a~0.5$^o~\times~0.5^o$ grid and assumed currents originate~110~km above the surface. When applied to different region, this altitude should be tested to ensure accurate results. The accuracy of SECS in modelling magnetic field variations between magnetic observatories in Ireland and the UK was evaluated by \citet{Campanya2019} using nine magnetometers, with respect to distance from the nearest magnetometer observatory. They highlighted that coherence and SNR (signal-to-noise ratio) fall off the further away the MT site was from the closest magnetic observatory, SNR of above 9 and coherence above 0.9 were observed for MT sites that are at less than 200 km from a magnetic observatory, which is the current situation in Ireland. However, the relative position of the site relative to the magnetometers and can also have an important bearing on the accuracy of the interpolation (i.e. an output site in between input sites will work well, but a site not between will not work as well). The presented nowcasting methodology is focused on only using magnetometers from Ireland (the three sites in Fig. \ref{fig:Mag_MT_Sites}), thus the performance of SECS interpolation was re-assessed for the current set-up. This was considered to be important due to the relative position of the MT sites towards the magnetic observatories, as in some cases SECS will work more like an extrapolation rather than an interpolation method. The assessment of the potential for SECS was validated using a) Measured time series at the MT sites (Figure \ref{fig:Mag_MT_Sites}), b) Modelled time series with differing numbers of magnetometers (nine, three and two), expanding on validation by \citet{Campanya2019} by accounting for relative position of output sites to magnetometers. 
    
    \subsection{Geoelectric Fields: MT-TF}\label{section:Electrics}
    Geoelectric fields are generated by the interaction of magnetic field variations originating from the magnetosphere and ionosphere with the electrically resistive Earth. The dominant layer  of the Earth, in terms of generating geoelectric fields, is the lithosphere, as it is the closest layer to the surface and is usually more resistive than the layers below. Varying magnetic fields ($B$) induce electric fields ($E$) according to Faraday’s law. 
    In the case of the Earth’s subsurface, geoelectric fields are dependent on a) the local strength of magnetic variations and b) the electrical resistivity properties of the Earth, expressed by a complex impedance tensor Z,  
    \begin{equation}\label{equation:Efield}
        E(\omega) = Z(\omega)\cdot B(\omega).
    \end{equation}
    When the impedance tensor $Z$ (Eq. \ref{equation:GalDist}), is known, the electric field can be modelled using the magnetic field, or vice-versa \citep{{Tikhonov1950}, {Cagniard1953}} with MT-TF (Eq. \ref{equation:Efield}). To retrieve the estimated electric field time series (Eq. \ref{equation:MT-TF}), an inverse Fast Fourier transform is performed: 
    
    \begin{equation}\label{equation:MT-TF}
        E_{est}(t) = FFT^{-1}[Z(\omega)\cdot FFT(B_{SECS}(t))].
    \end{equation}
    \mbox{}\\
     where $B_{SECS}$ is the magnetic field estimated after using a SECS interpolation. The complex impedance tensor Z, can be expressed by:
    \begin{equation}\label{equation:GalDist}
        Z(\omega) =
        \left[ {\begin{array}{cc}
        Z_{xx} & Z_{xy} \\
        Z_{yx} & Z_{yy} \\
        \end{array} } \right].
    \end{equation}
    \mbox{}\\
    Note that each component of $E~(E_{x}, E_{y})$, $B_{SECS}~ (B_x, B_y)$ and $Z$ are complex numbers. The MT-TF across Ireland were obtained from SWEMDI. The database contains MT-TF derived from simultaneously measured electric and magnetic field time series at 43 sites. The MT-TF were derived focusing on periods between 20 - 32,000~s, which are appropriate for modelling geoelectric fields caused by geomagnetic storms \citep{{Grawe2018},{Oyedukun2020},{Trichtchenko2021}}. Although MT-TF have been proved to provide accurate results when modelling geoelectric fields in Ireland during geomagnetic storms \citep{Campanya2019}, it is important to remember that this approach is limited by the plane-wave approximation, which is not always true during geomagnetic storms and can introduce errors to the modelled geoelectric fields \citep{ErnstJankowski}. 
  
    \subsection{Galvanic Distortion}\label{section:Galvanic_theory}
    The impedance tensor $Z$ can, in certain cases, be affected by local galvanic distortion. Galvanic distortions are real distortions in the regional electric field measurements caused by local, near surface inhomogeneities, commonly observed in MT surveys \citep{{Ledo1998}, {Rao2014}, {Delhaye2017}}. Galvanic distortion can affect the impedance tensor Z, and has a relevant impact on MT geophysical studies as it can make accurate modelling difficult to impossible, and, if ignored, can lead to erroneous conclusions \citep{jones_2012}. In terms of GICs studies, galvanic distortions may locally modify the geoelectric fields at a particular site, producing results not representative of the surrounding area between the site and the surrounding sites. 
    
    \paragraph{}
    
    Electric distortion is described by a real-valued, frequency independent tensor $C$ which relates the observed MT impedance tensor $Z_d$ to the regional MT impedance tensor $Z$ in the absence of distortion:
    \begin{equation}
        Z_d(\omega) = CZ(\omega),
    \end{equation}
    
    \mbox{}\\
    where $\omega$ denotes frequency dependence \citep{{Bahr1988},{GroomBailey}}. Algebraically, the distortion matrix mixes the complex-valued impedance components and therefore affects the amplitude and phase information of individual impedance components, which can lead to the aforementioned difficulties. The impedance tensor, Z, can be decomposed into a phase ($\Phi$, Eq. \ref{equation:phi}) and an amplitude tensor \citep{Neukirch2019}. The impedance phase tensor, the inverse imaginary part multiplied with the real part of the impedance, is shown to be unaffected by distortion \citep{Caldwell2004}:
    \begin{equation}\label{equation:phi}
        \Phi = Re(Z_d)^{-1}~Im(Z_d)=C^{-1}~Re(Z)^{-1}~C~Im(Z)= Re(Z)^{-1}~Im(Z).
    \end{equation}
    
    \mbox{}\\
    This has been used extensively to interpret MT data without most of the challenges pertained to galvanic distortion by disregarding the distorted amplitude tensor \citep{{Patro2012}, {Tietze2015}, {Bakker2015}, {Samrock2018}}). While the phase tensor only holds information about subsurface induction processes, the amplitude tensor describes galvanic and inductive effects that complement the phase tensor information \citep{Neukirch2019}. Both the amplitude and phase tensors describe the same physical induction processes in the subsurface and it has been hypothesised that they should reflect similarly the present subsurface geometry, i.e. electric strike, dimensionality and anisotropy, allowing computation of an optimal distortion matrix that maximises this geometric similarity \citep{Neukirch2020}. While empirical, this method of estimating present electric galvanic distortion has been shown to be effective. Here, we use this method to reduce the impact of galvanic distortion and compare results for nowcasting geoelectric fields with and without distortion correction.

\subsection{FFT Analysis for Nowcasting}\label{section:FFTanalysis}

Before computing the geoelectric fields, the input magnetics at MT sites were manipulated for nowcasting. FFT (Fast Fourier Transform) derived time-series suffer from issues at the edge of the time series (and hence at real time) due to the lack of frequency information past the edge of the time series. Zero-padding is the standard tool for improving the performance of the FFT close to the edge of a time series which involves adding an array of zeros to the end of the input signal, which reduces errors in the output when the FFT is applied. \citet{Kelbert2017} successfully implemented a padding approach to estimate geoelectric fields using an adapted form of zero-padding, where the last recorded values were used instead of zeroes (Fig. \ref{fig:padding_types}). Both padding approaches, and the combination of both, were analysed.

\paragraph{}

The performance of the FFT at real-time was evaluated by comparing padded near real-time (or nowcast) modelling of geoelectric fields with standard non real-time modelling of geoelectric fields (where electric fields time series do not suffer from edge issues). Several storms (Table \ref{table:storms}) were considered during the assessment. A single rectangular window function was used over other FFT window functions as this window worked best closer to real time (compared to a Hanning and Tukey windows), indicating frequency resolution is more important than spectral leakage.
    
    \begin{table}[H]

    \caption[]{Storms used to optimise the model together with their corresponding Kp, local K-index maxima in Ireland during each storm. Storms of varying strength were studied to see whether storm strength influenced the padding or correction curve. The magnetometer sites (see Fig. \ref{fig:Mag_MT_Sites}) used for the SECS interpolation are given in the right column.}
    \begin{center}
    \begin{tabular}{|c|c|c|c|l}
    \cline{1-4}
    \textbf{Storm} & \textbf{Local K Max$^{\mathrm{a}}$} & \textbf{Kp Max$^{\mathrm{b}}$} & \textbf{Local stations} &  \\ \cline{1-4}
    17-18 March 2015  & 7                          & 8-                    & BIR, VAL                &  \\ \cline{1-4}
    22-23 June 2015  & 8                          & 8+                    & BIR, VAL                &  \\ \cline{1-4}
    08-09 May 2016  & 6                          & 6+                    & BIR, VAL                &  \\ \cline{1-4}
    07-08 September 2017  & 7                          & 8+                    & ARM, BIR, VAL           &  \\ \cline{1-4}
    27-28 September 2017  & 5                          & 6+                    & ARM, BIR, VAL           &  \\ \cline{1-4}
    26-27 August 2018  & 7                         & 7+                    & ARM, BIR, VAL           &  \\ \cline{1-4}

    \end{tabular}
    \label{table:storms}
    \begin{list}{}{}
    \item[$^{\mathrm{a}}$] Derived from local magnetometers.
    \item[$^{\mathrm{b}}$] From German Research Centre for Geosciences (GFZ, \newline \url{https://www.gfz-potsdam.de/en/kp-index/}).
    \end{list}   
    \end{center}
    \end{table}
    
    Length and type/shape of the padding were evaluated to select the parameters that provide more accurate geoelectric fields. Potential parameters were analysed over several spatio-temporal scenarios, considering all 43 sites over Ireland (Fig. \ref{fig:Mag_MT_Sites}). Storms of varying strength (Table \ref{table:storms}) were chosen to see whether the model performed similarly for strong and weak storms. Real time electric field modelling, nowcasting, was compared to the modelled electric field (non real-time), to evaluate the performance of padded FFT. This was implemented as follows: 
    
    \begin{enumerate}
        \item The real-time/nowcast model is calculated between the timeframe [t - 48~hours : t], where t denotes real time (i.e. the last point in the magnetic field time series, the nowcast magnetic field). The non real-time synthetic model (standard model) is calculated between [t - 43~hours: t + 5~hours] (data past the defined real time for the non real-time model, as for the non real-time model, we have the future magnetic field time series).
        \item The points between [t - 5~hours: t] (at the edge for nowcast, away from the edge for standard) are compared between each model. 
    
        \item The relative error (Eq. \ref{eq:rel-error}), root-mean-squared error (RMS, Eq. \ref{eq:RMS}), signal-to-noise ratio (SNR, Eq. \ref{eq:SNR}) and coherence (Eq. \ref{eq:Coh}) were calculated to compare the performance at different delays from real time (between [t - 5~hours: t]) and are recorded. Here, we define delay ($\Delta{t}$) as the interval of time between real-time/now ($t$), and the time series at an earlier point ($T$), $\Delta{t}~=~t~-~T$. The purpose of calculating these metrics at each delay between [t - 5~hours: t] is to improve the accuracy not only at t, but previous times close to real time, to provide the most accurate results at each delay for the generation of a real-time movie.
        
        \item  Steps 1) to 3) were recalculated, as the time series was moved along a sliding window for the course of the storm.
        
        \item The mean of metrics from step 3) (relative error, root-mean-squared error (RMS), signal-to-noise ratio (SNR) and coherence, see Equation \ref{eq:meanRMS} in the Appendix) are calculated at each delay.

    \end{enumerate}

    \begin{figure}[H]
        \centering
        \includegraphics[scale=0.55]{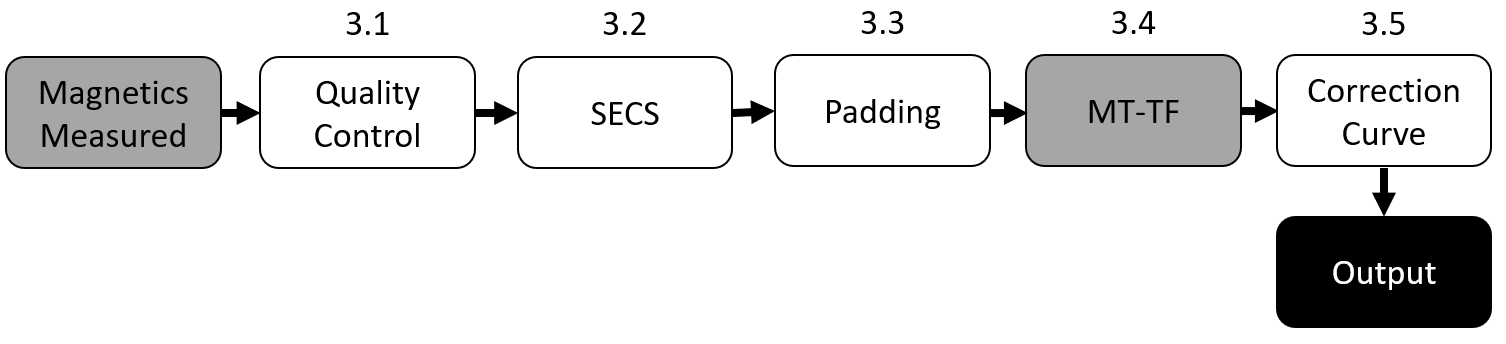}
        \caption{The main steps for computing geoelectric fields using our real-time method. White indicates the steps involved in preprocessing/manipulating the magnetic and electric field time series, grey indicates inputs, while black indicates the final electric field ouput. The method was calibrated using MT-TF corrected for galvanic distortion and validated against measured geoelectric time series, which were not corrected for galvanic distortion. Steps in the implementation are labeled based on their corresponding sections 3.1, 3.2,...
    }
        \label{fig:Workflow}
    \end{figure}

\section{Implementation and Validation}\label{section:Implementation}

    The nowcast approach, Figure \ref{fig:Workflow}, was implemented as follows:
    
    \paragraph{}
    
    \mbox{}\\
    3.1: The geomagnetic field was measured in real time with magnetometers, filtered for noise. (Section \ref{section:preprocesing}). 
    
    \mbox{}\\
    3.2: SECS are implemented to estimate magnetic field variations across Ireland (Section \ref{section:resultsSECS}).
    
    \mbox{}\\
    3.3: The magnetic time series were padded  and an FFT was performed to convert to the frequency domain (Section \ref{section:resultspadding}).
    
    \mbox{}\\
    3.4: Geoelectric fields were estimated using MT-TF (following the approach of \citet{Campanya2019}) with padded magnetic time series inputs (considering MT-TF corrected and non-corrected for galvanic distortion) and an inverse FFT converts back to the time domain (Section \ref{section:Geoelectric field calculation}).
    
    \mbox{}\\
    3.5: Amplitudes of the geoelectric field time series close to real time are multiplied by a correction curve (Section \ref{section:correction_curve}).

    \paragraph{}
    \mbox{}\\
     The nowcast model was then validated (Section \ref{section:validation}), the geoelectric fields were then visualised across the entire grid using a contour map (Section \ref{section:visualisation}) and the effect of using a galvanic distortion correction was examined (Section \ref{section:Galvanic_dist}).
    
    \subsection{Measurement , Preprocessing and Quality Control}\label{section:preprocesing}
    
    Magnetic field measurements were taken at individual magnetometer observatories (note, magnetic field data from the individual MT sites is not used here). Data for the last 48 hours (172,800~seconds) was used to ensure enough information about the largest periods of analysis (32,000~seconds, $>$~5 wave cycles) was retained (after testing we found 48 hours to be most suitable for our purposes). This data was linear detrended and then treated for any high frequency noise. A simple $dB/dt$ threshold was employed in real time to filter noise from moving ferrous sources (i.e., cars). This filter removes any signal above a 10~nT/sec $dB/dt$ threshold, which is sufficiently sharp and large to ensure no real signal is removed. Periods outside the range of interest (20~s -- 32,000~s) were not considered.

    \subsection{SECS }\label{section:resultsSECS}
    The potential of SECS approach to interpolate between the MagIE monitoring stations to MT sites (Fig. \ref{fig:Mag_MT_Sites}) was assessed by:

    \begin{enumerate}
        \item Comparing the SECS-derived magnetic fields with measurements at time of acquisition of MT sites.
        \item
        Quantifying the performance of SECS, comparing the results of using magnetometers from Ireland, UK, France, Belgium and Germany (the nearest 6 INTERMAG observatories) with results using only the available magnetometer data from Ireland, considering that 3 or 2 of the magnetometers are available. 

    \end{enumerate}

    The analysis was performed assuming that the three magnetometers are working, but also with the hypothetical case where the northern magnetometer stops working. SECS interpolations are more accurate during greater levels of magnetic activity, which is when we are more interested in the geoelectric time series. However, not all MT sites have data acquired at storm times, therefore we chose a magnetic activity of Kp4 as the minimum threshold for the site to be evaluated (extra analysis in Fig. \ref{fig:kp_cohsnr}). 
    
    \paragraph{}

\begin{figure}[H]
    \centering
    \includegraphics[scale=0.5]{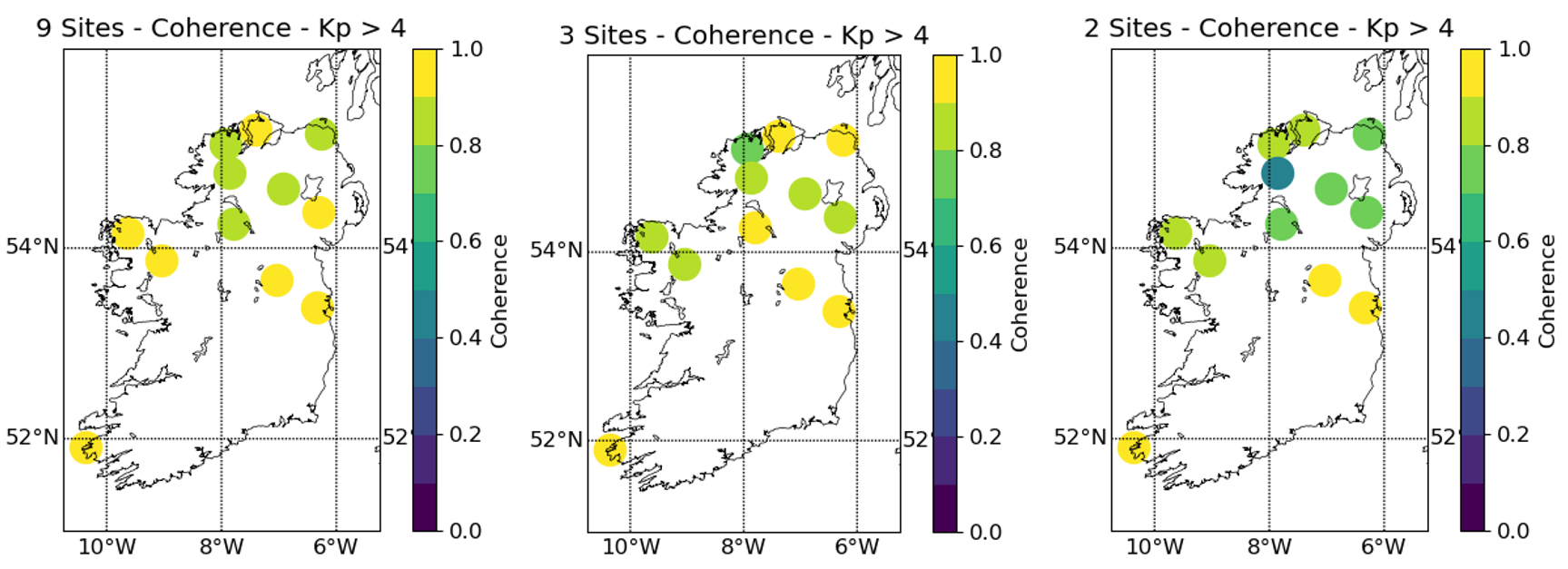}
    \includegraphics[scale=0.5]{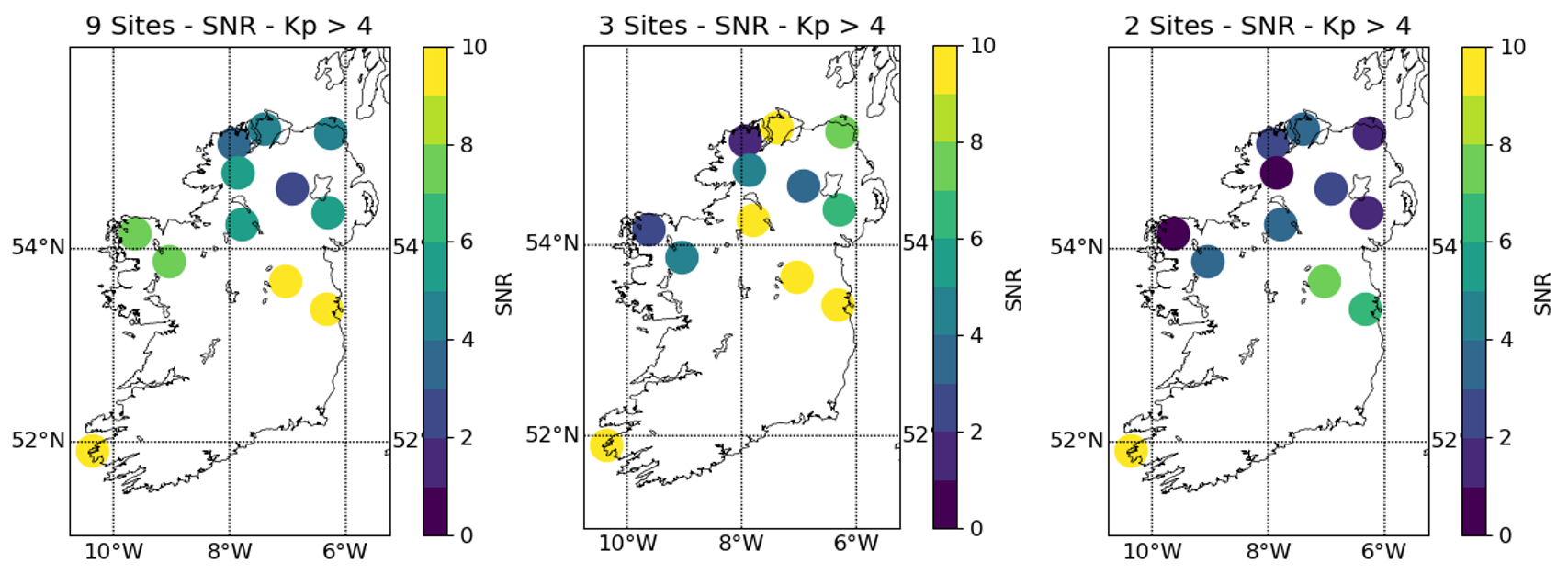}
    \caption{ The coherence (top) and SNR (bottom) between the measured magnetic field at MT sites and modelling magnetic fields (SECS interpolation with nine, three and two magnetometer sites) for all MT sites with available data. 
}
    \label{fig:coh_snr1}
\end{figure}

    Figure \ref{fig:coh_snr1} compares measured data at the time of calculation of the MT sites with SECS derived magnetic time series using magnetic time series from: a) nine magnetic observatories (the three from MagIE (ARM, BIR, VAL) with the nearest six INTERMAG observatories (HAD, ESK, LER, CLF, DOU, WNG), b) three magnetic observatories (VAL, BIR, ARM), and c) two observatories (VAL, BIR). The analysis was performed looking at coherence and SNR. Performance decreases with a lower number of magnetometers, in particular when going from three to two, missing the magnetometer in the north (Arm), but no significant difference were observed between using nine or three magnetometers, which suggest that the current configuration is reasonable, although it could improve if more data was available in real time. Sites with the strongest decrease in performance are the two sites further west. 
    
\begin{figure}[H]
    \centering
    \includegraphics[scale=0.6]{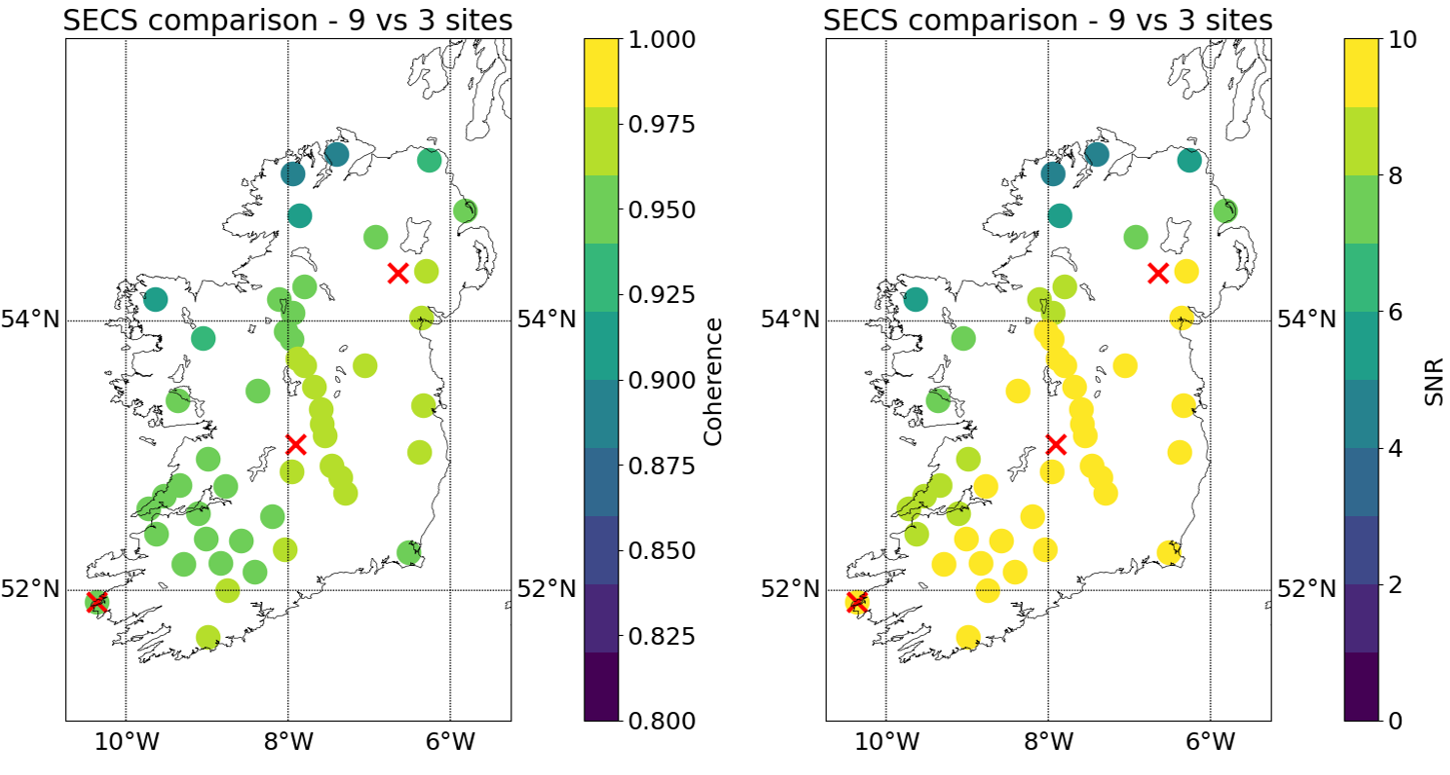}
    \includegraphics[scale=0.6]{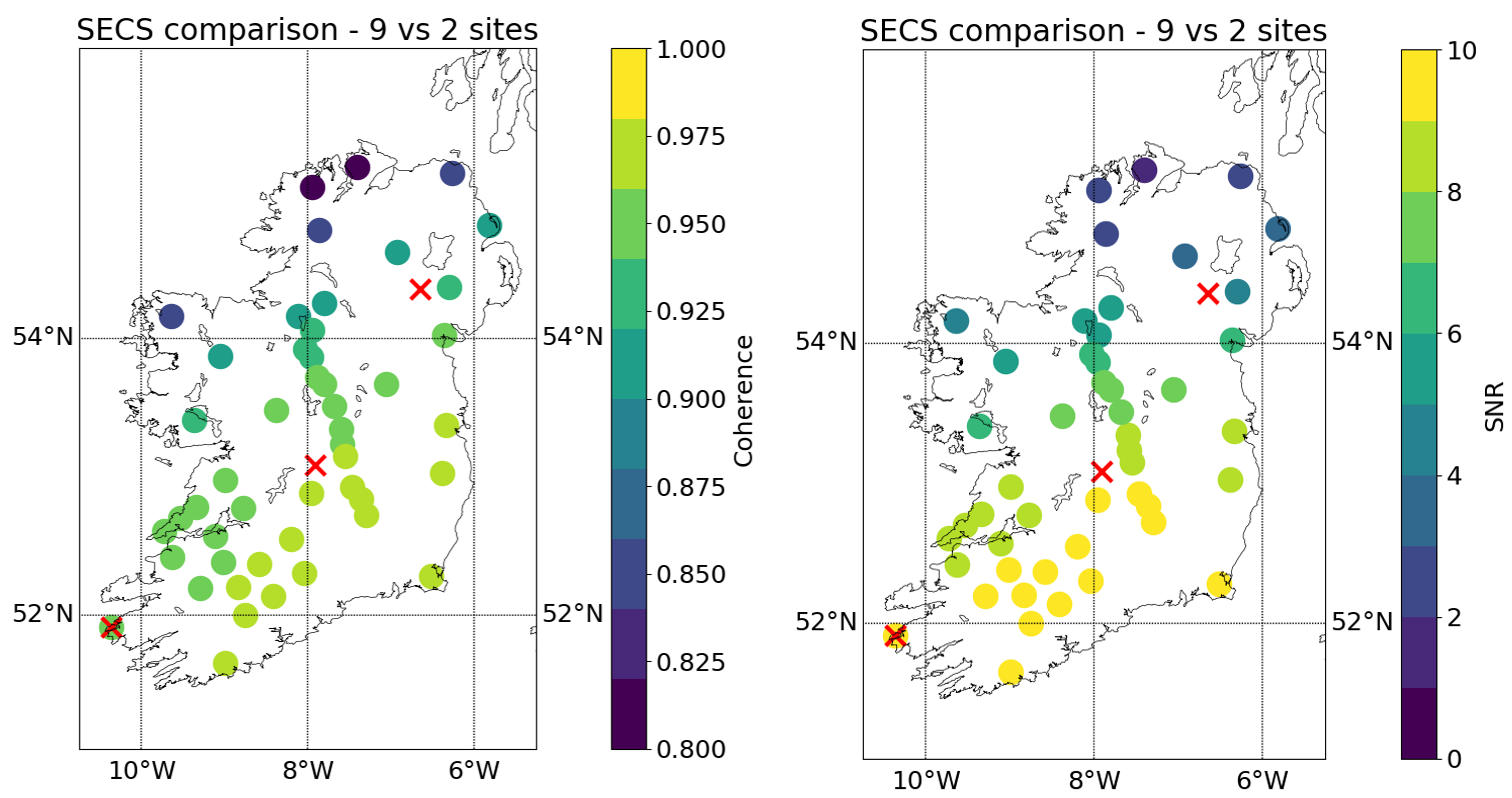}
    \caption{The coherence (left) and SNR (right) between modelled magnetic field using SECS with nine sites is compared to modelled magnetic field using SECS with three magnetometers (top) and two magnetometers (bottom) for the 7-8 September 2017 storm (Kp8). Positions of the MagIE magnetometers are denoted by red crosses.}
    \label{fig:coh_snr2}
\end{figure}

    \paragraph{}
    The impact of using two or three magnetometers was also evaluated by comparing with interpolating with nine magnetometers (Fig \ref{fig:coh_snr2}). The performance of the SECS method decreases  . The performance of the SECS method decreases when using only three sites, however it still performs well with high coherence ($>$ 0.875) and SNR ($>$ 4). The performance of the SECS method decreases significantly for northernmost sites, when the ARM magnetometer is removed, with high coherence ($>$ 0.8)  but low SNR (0 - 4). This highlights the importance of using the ARM magnetometer for accurate magnetic fields at northern sites. Despite the lack of a magnetometer in the south-east, these sites still perform well. This is likely due to the magnetometers retaining the regional effects, with weak local effects in the south-east, due to the lack of strong influence from the auroral electrojet, as opposed to the north-west, which is more strongly influenced.

    \subsection{Padding}\label{section:resultspadding}    
    
    To optimise padding, the performance metrics (relative error, SNR, Coherence and RMS) were analysed comparing the a padded electric field time series at real-time to an unpadded time series away from real-time (see Section \ref{section:FFTanalysis} for the method used). Electric fields below 2~mV/km were discarded for this analysis to help avoid division/multiplication by 0 for SNR and relative errors. Signals below 2~mV/km also pose a negligible risk in comparison to larger electric fields. 
    \paragraph{}

\begin{figure}[H]
        \centering
        \includegraphics[scale=0.6]{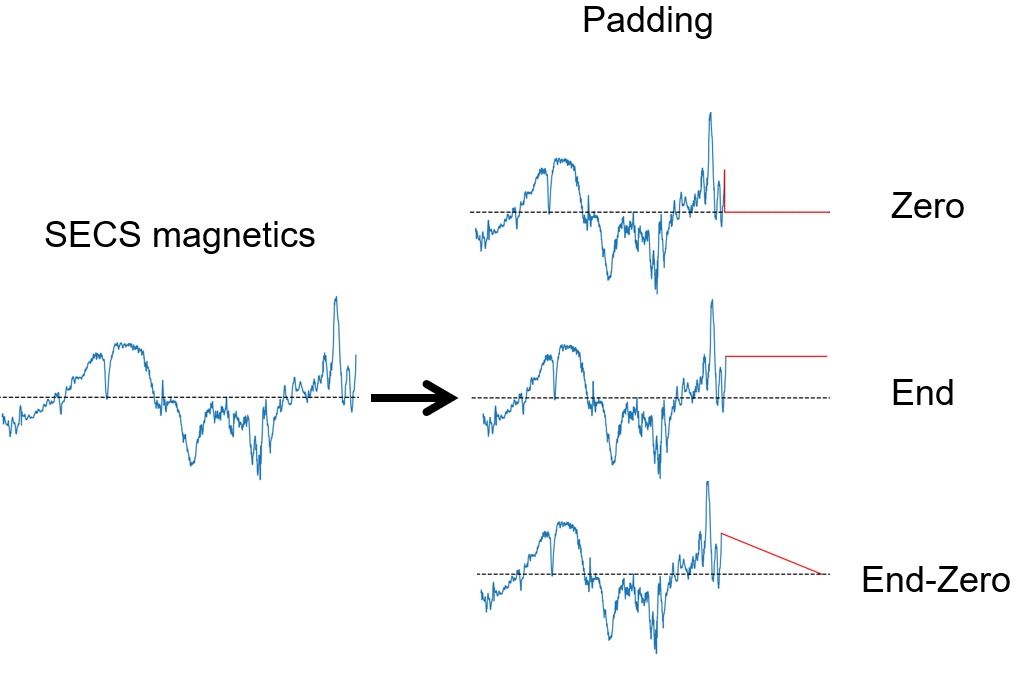}
        \caption{An example of the different padding types applied to the magnetic field time series, zero-padding, end-padding and end-zero padding.}
        \label{fig:padding_types}
    \end{figure}

    The best length of the padding was evaluated by using lengths between 0 and 180 min with 5 min intervals. The types of padding considered included zero padding, end padding and end-zero padding (Fig. \ref{fig:padding_types}). The best overall performance was obtained with a length of 105 min using either end padding or end-zero padding. The ratio of pad to time series (48 hours, 2880 min) is 0.036.  
    \paragraph{}

\begin{figure}[H]
        \centering
        \includegraphics[scale=0.45]{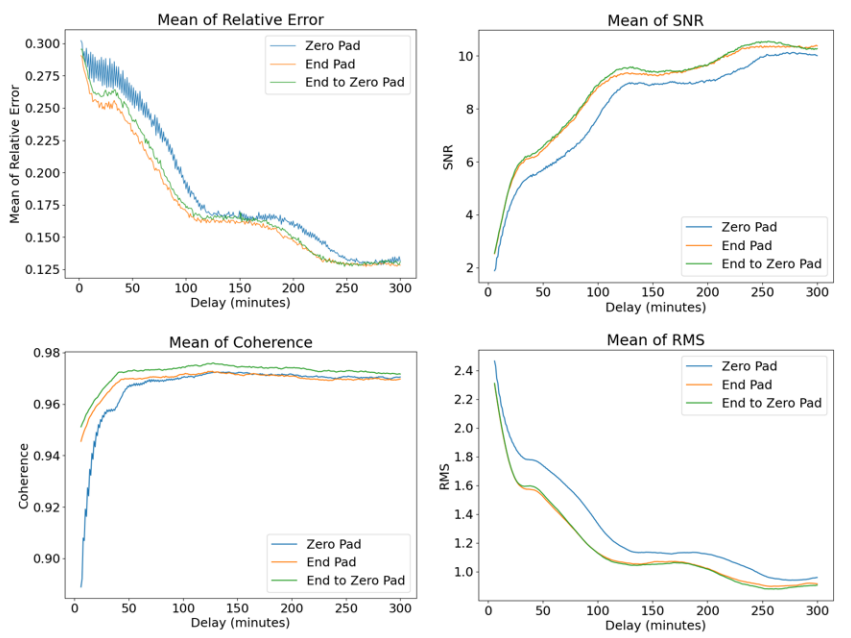}
        \caption{A comparison of the performance of “end-zero padding”, “end padding” and “zero padding”.}
        \label{fig:padding1}
    \end{figure}

\begin{figure}[H]
         \centering
         \includegraphics[scale=0.45]{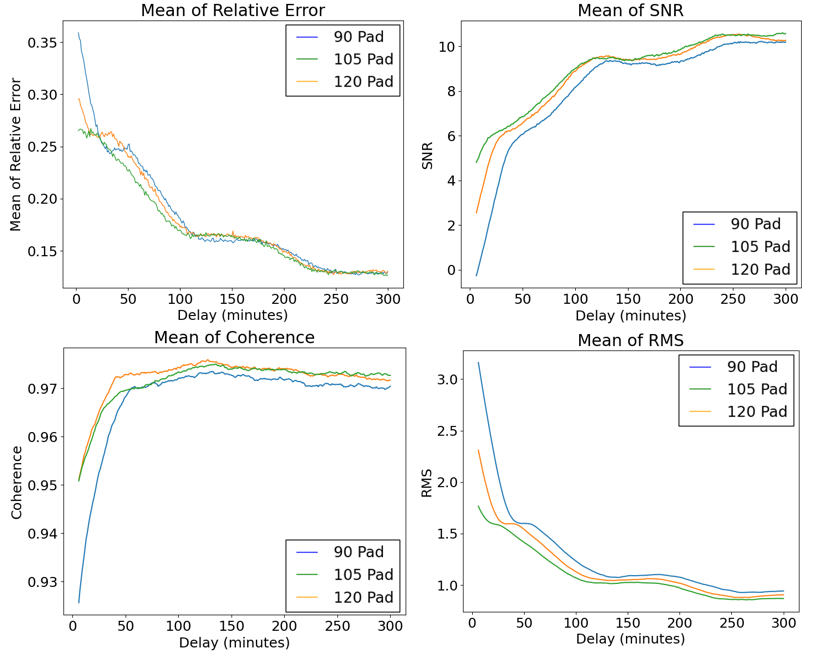}
         \caption{A comparison of the performance of end-zero padding with different lengths of padding (in minutes). }
         \label{fig:padding2}
     \end{figure}

    Firstly, padding type was analysed. Figure \ref{fig:padding1} compares the performance of the FFT with three different types of padding (using 120 min of padding, an estimate). End and end-zero padding outperform zero padding in each metric. The difference between end and end-zero padding is more subtle. Both perform at a similar level in each metric. The most significant difference between the two is in coherence, where end-zero padding is marginally better across all delays. For this reason, we chose to continue with end-zero padding as the padding method. 
    \paragraph{}
    Figure \ref{fig:padding2} expands on the analysis in Figure \ref{fig:padding1} by comparing the performance of end-zero padding using three different padding lengths. If the padding length was set too small, the coherence was not accurately re-produced by the model. If the padding was set too large, the SNR and RMS worsened (although coherence improved marginally). Note, despite the latter three storms in Table \ref{table:storms} missing ARM data for the interpolation, this had little impact on the performance metrics. This is likely due to the same input MT-TF being compared, so while the magnetic field will be more erroneous, this has little effect on the comparison between different padding types.  
    
    
    \subsection{MT-TF -  Geoelectric Field Calculation}\label{section:Geoelectric field calculation}
\begin{figure}[H]
        \centering
        \includegraphics[scale=0.6]{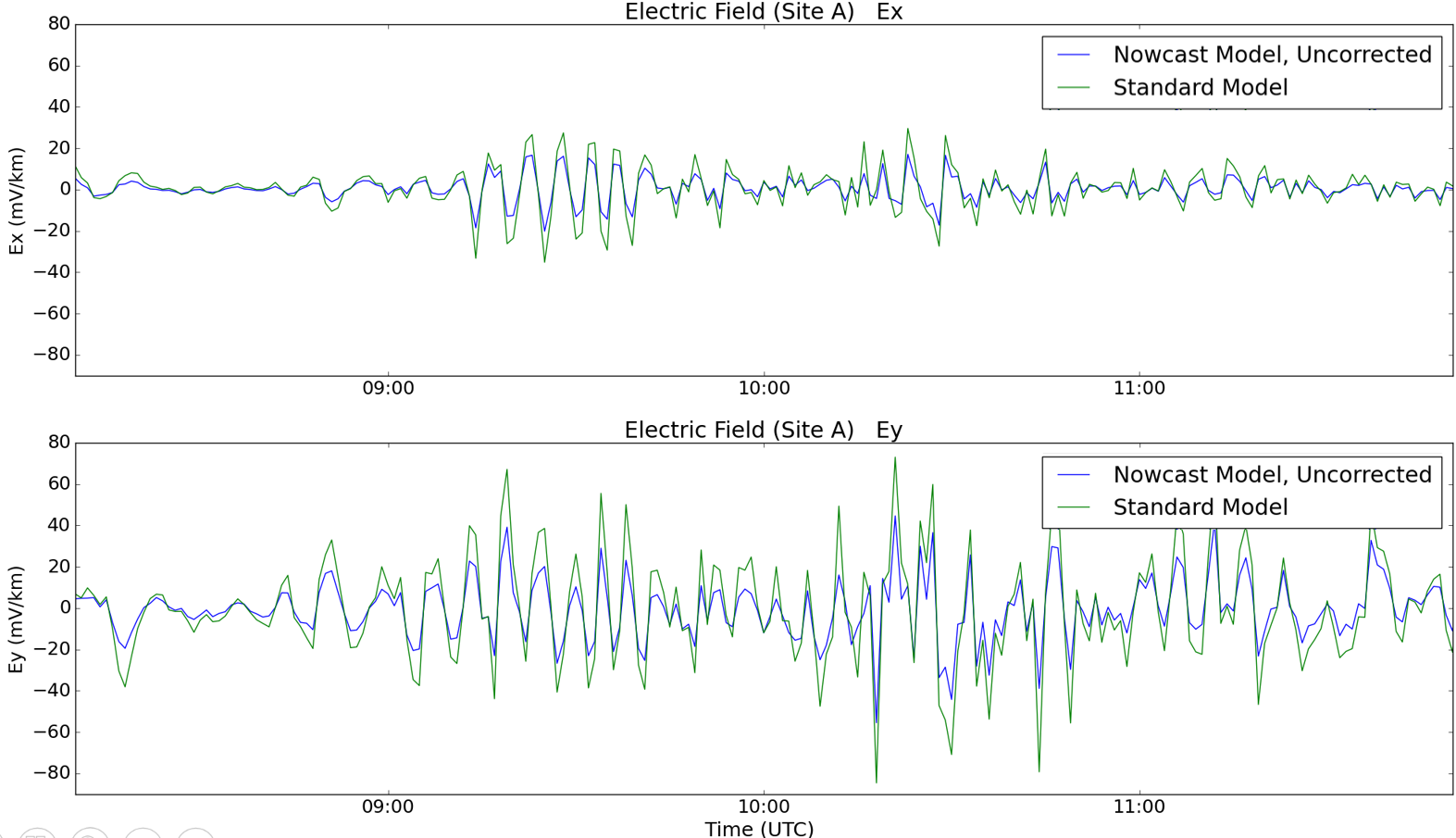}
        \caption{An example of the modelled time series of the electric field at a 1-minute delay (nowcast) plotted against a time series of the standard non real-time model for $Ex$ (top) and $Ey$ (bottom) for the 30 October 2003 for Site A (Fig. \ref{fig:Mag_MT_Sites}). }
        \label{fig:ModelledE}
    \end{figure}

     Geoelectric fields were modelled across Ireland using the MT-TF approach for all sites from SWEMDI (Figure \ref{fig:Mag_MT_Sites}).
     Padded magnetic field time series were converted to to frequency domain using an FFT, and  then the geoelectric field was  calculated at each site using Equation \ref{equation:MT-TF}. For the implementation, we focused on sites with corrected galvanic distortion, and kept the sites not corrected for galvanic distortion for the validation. Note that despite the use of padding, the 1~minute delayed/nowcast geoelectric field at this point underestimates peaks when compared with standard modelling (non-real time), but retains coherent structure (Fig \ref{fig:ModelledE}). This underestimation is due to the inability of the FFT to reproduce all of the long period information at the edge of the time series. However, short periods are retained and give spatial coherence to the results. In the following section we discuss how we attempt to account for this effect.
    
    \subsection{Correction Curve}\label{section:correction_curve}

\begin{figure}[H]
        \centering
        \includegraphics[scale=0.4]{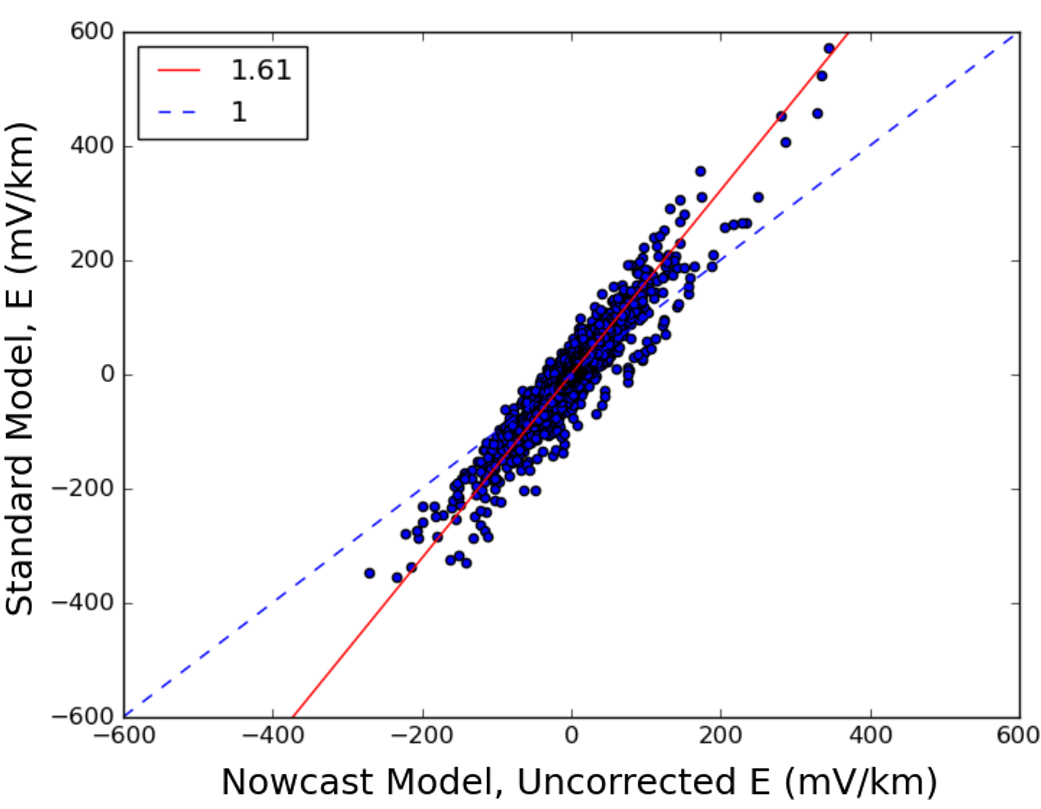}
        \caption{The amplitude of the standard electric field model (both $Ex$ and $Ey$) are plotted against the corresponding 1-minute delayed from real time (nowcast) electric fields, for all sites in Figure \ref{fig:Mag_MT_Sites} and storms in Table \ref{table:storms}. The ideal correction factor (in red) is the best fit slope between the two models and is compared to a slope of unity (in blue).}
        \label{fig:Corr_factor}
    \end{figure}

    A correction curve was considered to correct the underestimation of the nowcast geoelectric fields. The correction curve applies a scalar correction factor on the amplitude of the nowcast geoelectric fields for each time delay. This correction factor has been optimised for, by comparing the real-time and non real-time models (explanation of real-time vs non real-time in Section \ref{section:FFTanalysis}) for 5 storms (Table \ref{table:storms}) and for all MT sites, by calculating the average ratio (example at 1-minute delay, Fig. \ref{fig:Corr_factor}).

    \begin{figure}[H]
        \centering
        \includegraphics[scale=0.55]{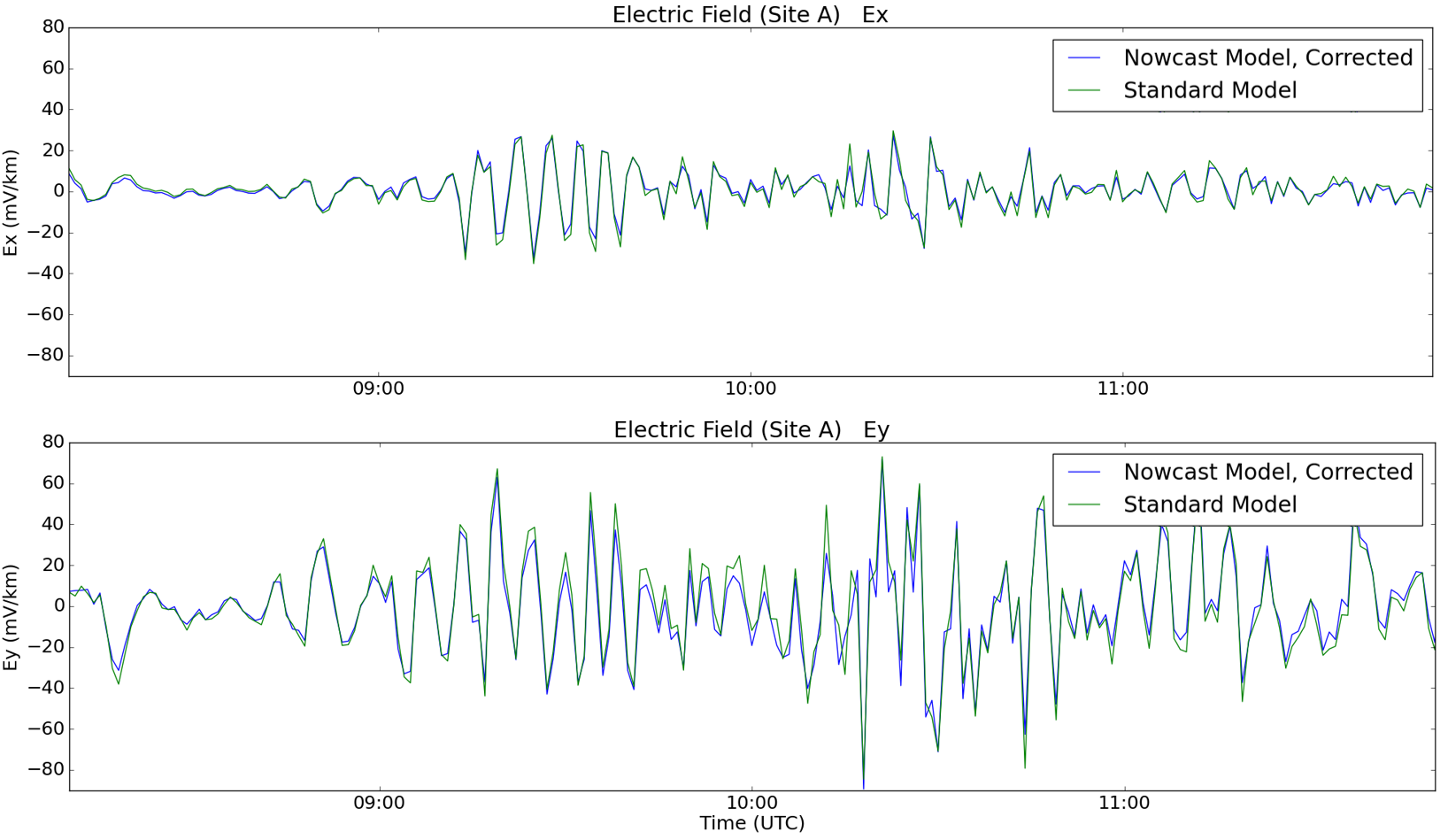}
        \caption{The corrected time series of the electric field at a 1-minute delay (nowcast) against the standard model for $Ex$ (top) and $Ey$ (bottom) from Figure \ref{fig:ModelledE}, with a pre-optimised correction factor of 1.61 (Fig. \ref{fig:Corr_factor}). }
        \label{fig:Model_E_corr}
    \end{figure}
    
\begin{figure}[H]
        \centering
        \includegraphics[scale=0.5]{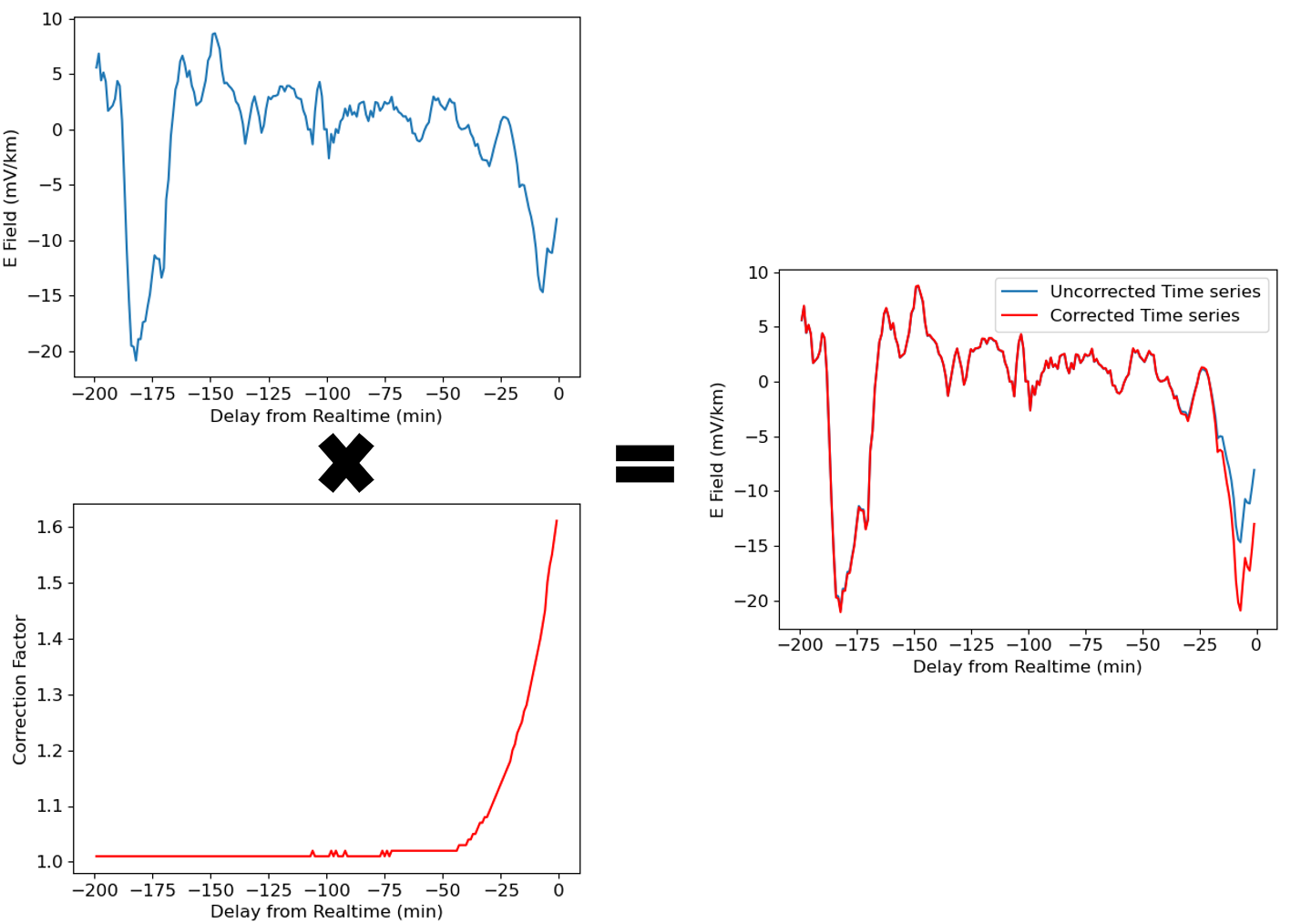}
        \caption{An example of the change in the time series near real time when the electric field time series is multiplied by the correction curve. The uncorrected time series (top left) are multiplied by a correction (bottom left), to give the corrected time series (right).}
        \label{fig:corr_applied}
    \end{figure}

\begin{figure}[H]
        \centering
        \includegraphics[scale=0.7]{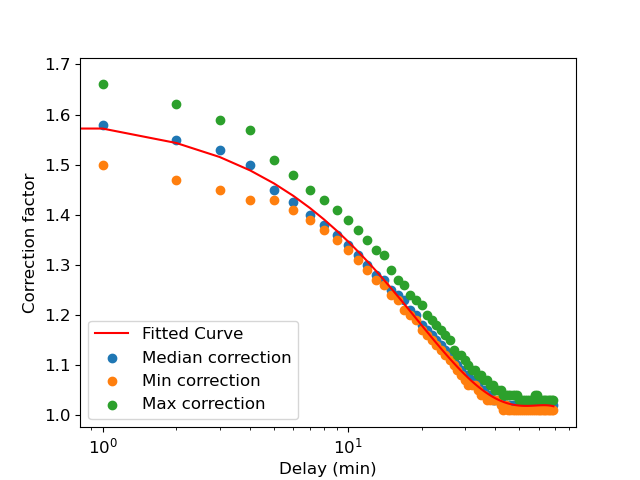}
        \caption{The median correction factor calculated at each time delay (the correction curve). The maximum (green) and minimum (orange) extent of the curve at each delay are also marked.}
        \label{fig:Correct_Curve2}
    \end{figure}

      A performance score (Eq. \ref{Eq:perf}) was used to optimise the model's correction factor. The correction curve was defined by considering correction factors between 0.5 and 2 with 0.01 increments, constraining median, min and max correction factors for each time delay between 1 and 200 min. Figure \ref{fig:corr_applied} shows a graphical summary of the role of the correction curve. Figure \ref{fig:Correct_Curve2} shows the median, minimum and maximum values of the correction curve for each time delay between each site and storm. 

    The correction curve is dependant on the length of time series used (48 hours in this case) and can be given by the conditional equation:

\begin{equation}
    f(t) \approx
    \begin{cases}
        - (3.0 \times 10^{-6}t^{3}) + (5.3 \times 10^{-4}t^{2}) -(3.0 \times 10^{-2}t) +1.61 & {\text{if }  x/n \leq 0.0024}\\
        1.00 & {\text{if }  x/n > 0.0024}
    \end{cases}
\end{equation}

\mbox{}\\
where n is the length of the time series used, t is the time delay, 0.0024 is derived from the length of the time series 2880, divided by delay 70, the edge of where a correction is needed (see Figure \ref{fig:Correct_Curve2}). As the amplitude of the correction factor reduces, it gets further than real time, which is expected, as more longer period variations are present closer to the centre of the time series, so the underestimation becomes less and less. At approximately 70 minutes the correction factor approaches unity, indicating that any loss in long period information here and at larger delays is negligible.

\paragraph{}

Applying the correction at each delay is performed to construct an nowcast movie of geoelectric fields (to visualise how the electric fields develop up until real time). The correction curve multiplies the electric field time series by a correction factor at each time delay, close to real time, to account for lost longer period magnetic variations (i.e. the electric field time series is multiplied by 1.61 at 1 minute, 1.58 at 2 mins, etc.). Rather than re-calculate the correction at each delay individually, the curve can be used to account for the error at each delay. This significantly reduces computation time. 
    
    \subsection{Validation}\label{section:validation}
    
\begin{figure}[H]
    \centering
    \includegraphics[scale=0.60]{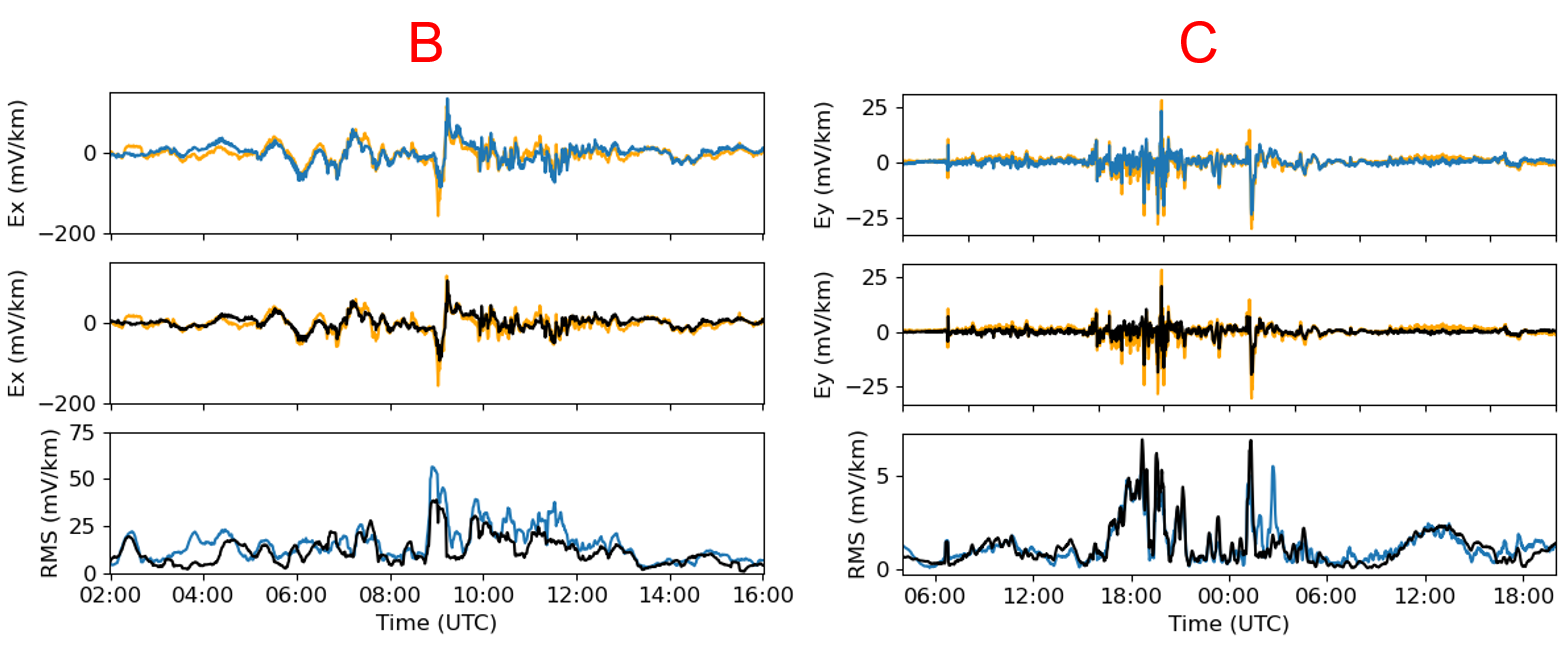}
    \includegraphics[scale=0.60]{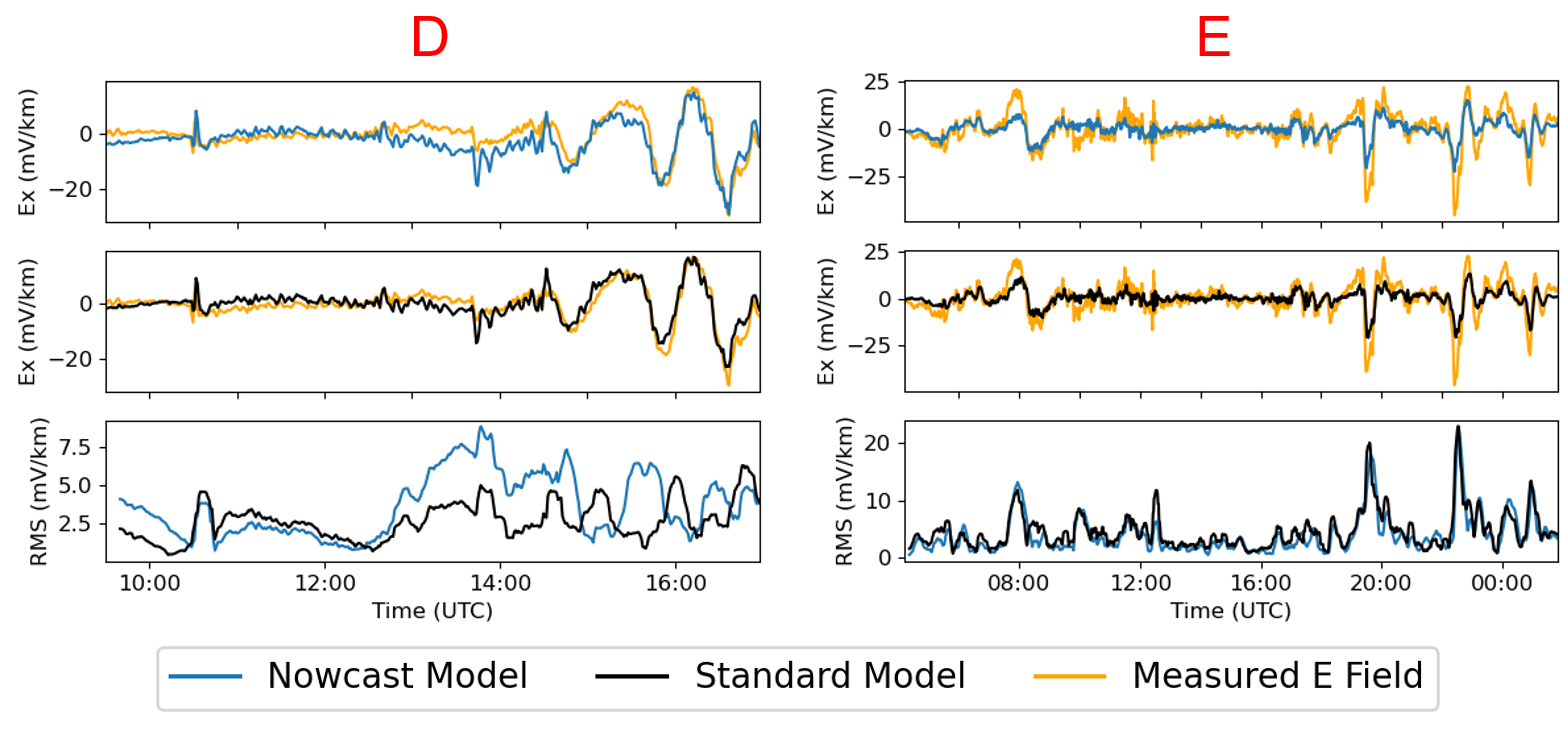}
    \caption{The measured and modelled E are compared  for four example sites (Fig. \ref{fig:Mag_MT_Sites}), (top left) 17-18 March 2015 (Kp8) for site B, (top right) 07 November 2021 (Kp7) for site C, (bottom left) 21 November 2017 (Kp5) for site D and (bottom right) 05 May 2018 (Kp5) for site E. The top subplot compares the measured electric field (gold) to the nowcast model at a 1-minute delay (blue), the middle subplot the measured electric field versus the standard model (black), with the bottom comparing the RMS (10~minute bins) between the models and the measured data.}.
    \label{fig:Geo_Comp_2}
    \end{figure}

    The presented approach for nowcasting geoelectric fields was validated with clean measured electric field data (band pass filtered between 20 - 32,000s) from the SWEMDI database \citep{Campanya2018}. During validation galvanic distortion was not corrected as it is present in the measured electric field data. Four measurement sites were used for validation \ref{table:storms2}. Validation results are presented in Figure \ref{fig:Geo_Comp_2} and summarised in Table \ref{table:validations}. Note that data measured at these sites during the selected storms were not used for implementing/optimising/calibrating the system for nowcasting geoelectric fields.

\begin{table}[H]

    \caption[]{The storms used to validate the model with their corresponding Kp and local K-index maxima in Ireland during each storm. Storms of varying strength were studied to see whether storm strength influenced the padding or correction curve. The magnetometer sites (see Fig. \ref{fig:Mag_MT_Sites}) used for the SEC interpolation are given in the right column.}
    \begin{center}
    \begin{tabular}{|c|c|c|c|c|l}
    \cline{1-5}
    \textbf{Storm} & \textbf{Local K Max$^{\mathrm{a}}$} & \textbf{Kp max$^{\mathrm{b}}$} & \textbf{Local stations} & \textbf{MT Site} \\ \cline{1-5}
    17-18 March 2015  & 7                          & 8-                    & BIR, VAL & B               &  \\ \cline{1-5}
    22 November 2021  & 6                          & 8-                    & ARM, BIR, VAL  & C               &  \\ \cline{1-5}
    21 November 2017  & 5                          & 5                    & ARM, BIR, VAL   & D             &  \\ \cline{1-5}
    07-08 September 2017  & 5                          & 6-                    & ARM, BIR, VAL & E          &  \\ \cline{1-5}

    \end{tabular}
    \label{table:storms2}
    \begin{list}{}{}
    \item[$^{\mathrm{a}}$] Derived from local magnetometers.
    \item[$^{\mathrm{b}}$] From German Research Centre for Geosciences (GFZ, \newline \url{https://www.gfz-potsdam.de/en/kp-index/}).
    \end{list}   
    \end{center}
    \end{table}

    \begin{table}[H]

    \caption[]{The storms used to validate the model with their corresponding Kp index value and max local K-index in Ireland during each storm. Storms of varying strength were studied to see whether storm strength influenced the padding or correction curve. The magnetometer sites (see Fig. \ref{fig:Mag_MT_Sites}) used for the SEC interpolation are given in the right column.}
    \begin{center}
    \begin{tabular}{|c|c|c|c|c|l}
    \cline{1-5}
    \textbf{MT Site} & \textbf{Model} & \textbf{Coherence} & \textbf{SNR} & \textbf{RMS} & \\ \cline{1-5}
    B  & Standard                          & 0.80                    & 4.3 & 1.3             &  \\ \cline{1-5}
    B  & Nowcast                          & 0.75                    & 3.2   & 1.7            &  \\ \cline{1-5}
    C  & Standard                          & 0.89                    & 5.4   & 17             &  \\ \cline{1-5}
    C  & Nowcast                          & 0.81                    & 6.5     & 15      &  \\ \cline{1-5}
    D  & Standard                          & 0.90                    & 7.0    & 3.9            &  \\ \cline{1-5}
    D  & Nowcast                          & 0.84                    & 4.0    & 3.8            &  \\ \cline{1-5}
    E  & Standard                          & 0.83                    & 3.8    & 4.0            &  \\ \cline{1-5}
    E  & Nowcast                          & 0.89                    & 4.5     & 3.3      &  \\ \cline{1-5}
    \end{tabular}
    \label{table:validations}
    \end{center}
    \end{table}

    \paragraph{}
    Each site in Figure \ref{fig:Geo_Comp_2} is ordered in terms of distance to magnetometer observatories (Figure \ref{fig:Mag_MT_Sites}, B is the closest, E is the furthest). In terms of overall accuracy compared to measured data, site C and D perform best in terms of the metrics (Table \ref{table:validations}). Site B is more erroneous as the MT  site is located at the coast, with extra noise present in the electric field due to tidal effects. For site E, while coherent shape is accurate, the electric field is underestimated, due to the underestimation in prediction of magnetic fields with SECS, related to distance and position relative to the magnetometers (Figures \ref{fig:coh_snr1},  \ref{fig:coh_snr2}). Comparing between the standard model and nowcast model, a small decrease in accuracy (drop in coherence and SNR), is present in general, due to the loss of high frequency information near real time, when the FFT is performed. However, some discrepancy is observed. At some sites (C, E), SNR is higher for the nowcast, as the nowcast overestimates the standard model, which itself underestimates the measured data. And at site E, coherence of the nowcast is greater, indicating that there may have been noise present at longer periods, for measurements of this site.

    \subsection{Visualisation}\label{section:visualisation}
    
    \paragraph{}
    To plot the real-time electric fields across the entire grid, a contour map was used. The map uses modelled electric field, amplitude and direction,  for each MT site in Figure \ref{fig:Mag_MT_Sites}. The electric field is then estimated across 10~km $\times$ 10~km ($0.1^o~\times$ $0.1^o$) grid encompassing the island of Ireland . The direction of the electric field at each impedance tensor site is also displayed. A cubic spline interpolation is then performed to create the contour and a mask is then applied to the sea.
    
    \begin{figure}[H]
        \centering
        \includegraphics[scale=0.55]{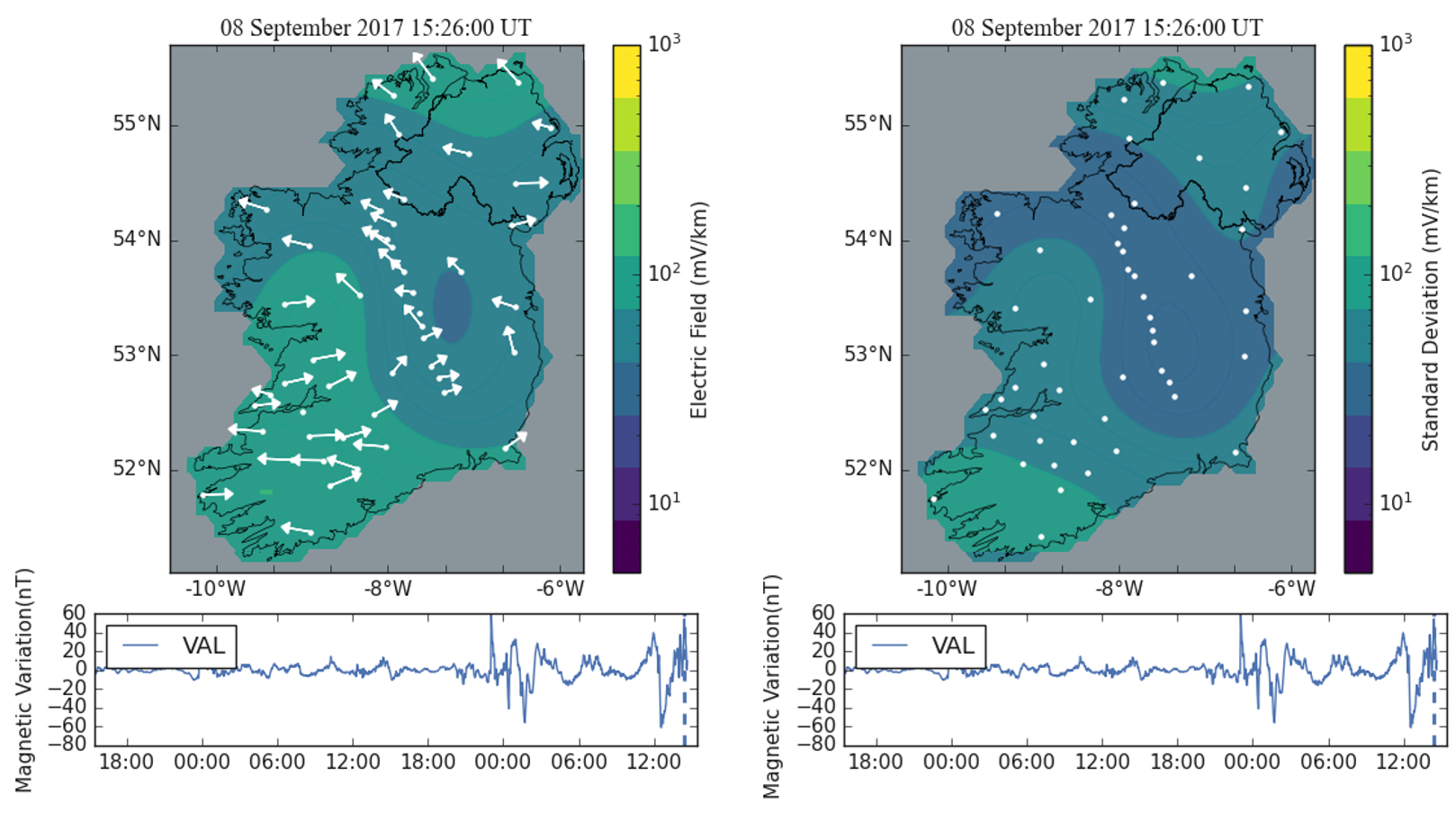}
        \caption{An example of the real-time model operating for the 8-9 September 2017 storm. The model treats data like it is in real time. The nowcast electric fields were plotted across Ireland at the top, with colour indicating magnitude and the arrows indicating direction on the left and the standard deviation error of the model on the right. A cubic interpolation used to estimate the electric fields between impedance tensor sites.  The horizontal (H) component of the Valentia magnetometer is on the bottom to illustrate the changing magnetic field. (An animation of the storm is attached in the supplementary materials).}
        \label{fig:Visualisation}
    \end{figure}
    
    \paragraph{}
    
    The real-time map of geoelectric fields is available at the MagIE website (\url{www.magie.ie/geoelectrics}). The standard deviation map considers the range of error in the geoelectric field measurements and SECS interpolation, they are proportional to amplitude and usually vary depending on the quality of the MT data and distance from nearest magnetic observatory.

    \subsection{Galvanic distortion}\label{section:Galvanic_dist}
    The regional effects of the model with and without galvanic distortion were examined:
    
    \begin{figure}[H]
        \centering
        \includegraphics[scale=0.60]{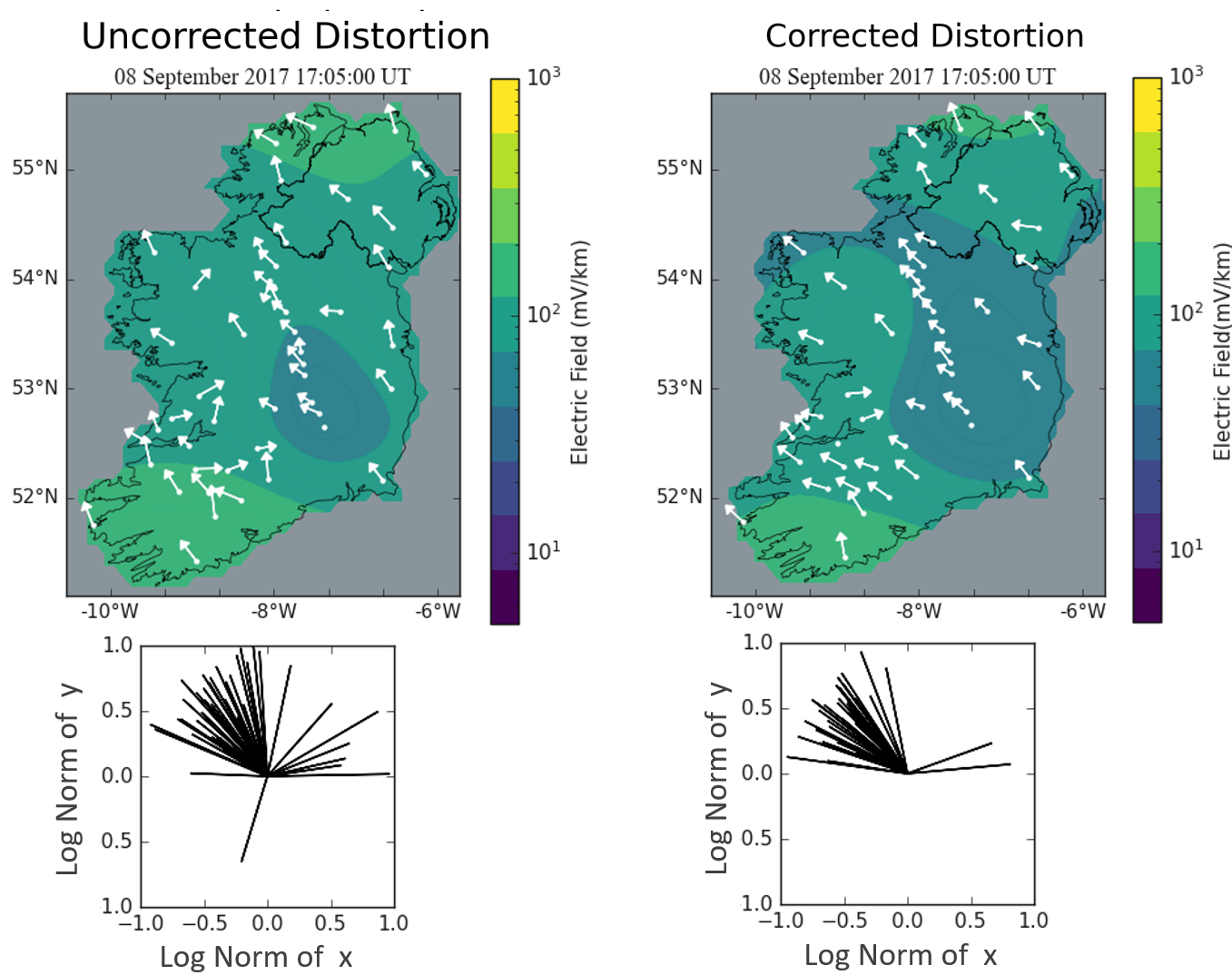}
        \caption{(Top) The model with undistorted electric field (left) is compared to the model with distorted electric field (right). (Bottom) The normalised direction of the vectors above are compared for significantly strong ($>~5~mV/km$) electric fields at the same origin. }
        \label{fig:Visual_direction}
    \end{figure}

    \begin{figure}[H]
        \centering
        \includegraphics[scale=0.4]{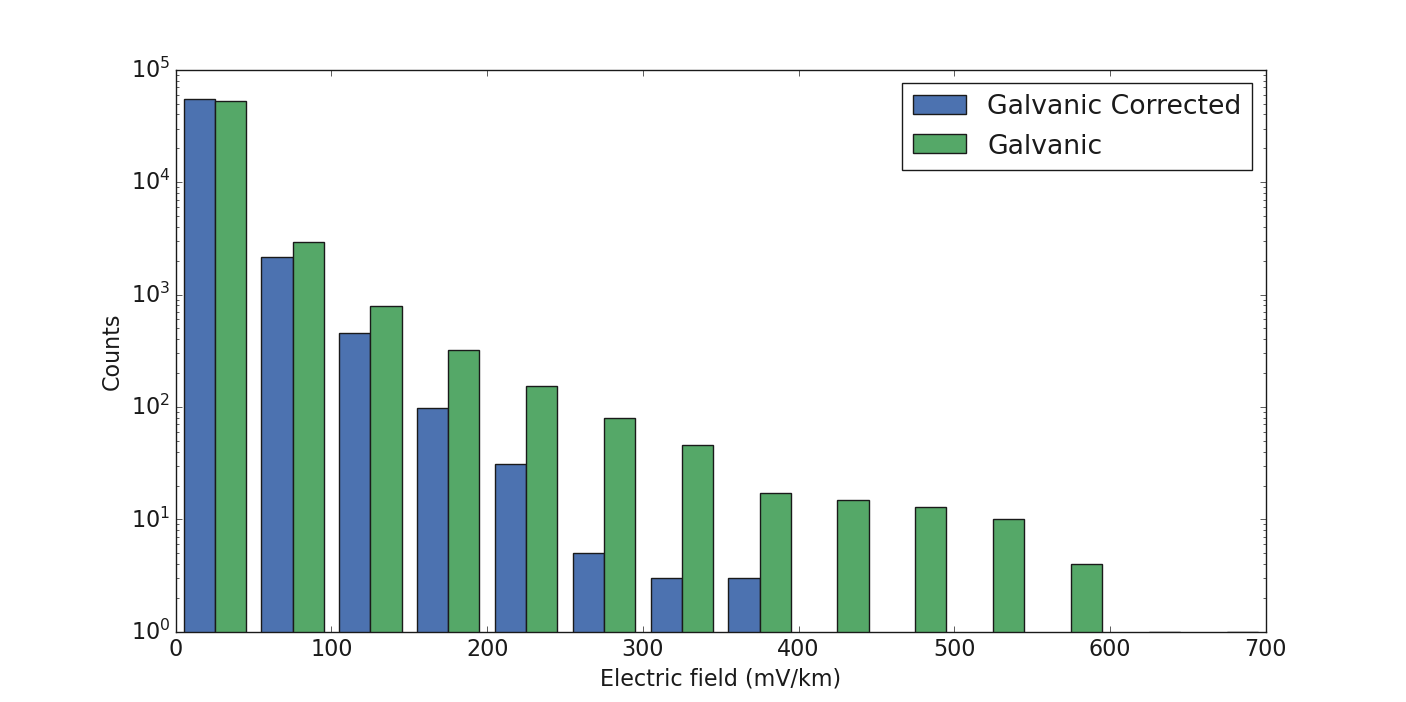}
        \caption{The frequency of amplitudes ($50~mV/km$ bins) at each site, for the entire storm in Figure \ref{fig:Visual_direction} are compared using a bar chart with undistorted electric fields (blue) and distorted electric fields (green). }
        \label{fig:Gal_hist}
    \end{figure}
    
    \paragraph{}
    Galvanic distortion was corrected for using the method from \citet{Neukirch2020} (explained in Section \ref{section:Electrics}). Two main differences between the models are present: 1) the local direction of the electric field between sites is less homogeneous with galvanic distortion (Fig. \ref{fig:Visual_direction}), 2) larger electric fields are generally observed when no correction is applied for galvanic distortion (Fig. \ref{fig:Gal_hist}). One would also expect there to be more volatile peaks in the model with galvanic distortion in Figure \ref{fig:Visual_direction},  however the cubic spline interpolation cleanses this effect. Locally in Figure \ref{fig:Visual_direction}, some patterns are also apparent, a) sites in the centre of Ireland have comparably weak electric fields compared to the sites closer to the coasts and the South-West, which could be explained by underlying lithospheric features in their respective regions.  b) galvanic distortions appear to have the greatest effect near the coast, often being orders of magnitude stronger than the galvanic corrected electric field, which matches results from previous findings \citep{jones_2012} .

    \section{Discussion}\label{section:discussion}
    A framework has been developed for nowcasting geoelectric fields in Ireland, caused by space weather events.  The framework was calibrated using modelled geoelectric fields, and validated against measured geoelectric fields, producing reasonable, albeit slightly worse results, when compared to the standard non real-time model, which is to be expected given the loss in frequency information. The framework also considers correction for galvanic distortion of the MT-TF and evaluated its impact on modelled geoelectric fields across the region. Spatial resolution is comparable to other national scale attempts to model geoelectric fields (i.e. NOAA for the US \footnote{NOAA Geoelectric Field 1-Minute, available at \url{https://www.swpc.noaa.gov/products/geoelectric-field-1-minute} as of 12/01/2023}), with no site in Ireland less than 200~km from magnetic observatories, and MT-TF at around ~40~km distance between sites on average. 
    \paragraph{}
    
    This framework was built in a way that it could be implemented outside Ireland, providing that real time magnetic field variations and MT-TF for the area of study are available. However, it is important to highlight that the framework was calibrated using data from Ireland, that no major changes are to be expected, it is recommended that the calibration/optimisation process is repeated if implemented in a new region. The framework represents a key step for space weather and GIC studies in Ireland provides for the first time geoelectric field data in near real time for the whole island. Data will be stored at \url{www.magie.ie} where it can be used to assess the impact of past events.
    
    \subsection{Nowcasting Geoelectric Fields}\label{section:Discussion-Geoelectrics}

    The nowcast geoelectric fields method used the FFT-based method of \citet{Kelbert2017} as its foundation and built upon it. Firstly, the optimisation of the method used by \citet{Kelbert2017} was elaborated upon, including a more detailed description of how to calibrate an ideal padding length. In addition, a correction curve was added to account for the underestimation at modelling geoelectric fields near real time. The considered correction curve approach is similar to the method used in \citet{SimpsonBahr}, where they correct the individual MT-TF, instead of correcting the electric field time series. However, we found the correction curve to be more practical for our purposes (i.e. for a real-time movie), when correcting multiple delays from real time, as separate MT-TF would need to be made for each separate delay, which would reduce computation time and would be tedious to perform for many MT sites. The correction curve improves the RMS of the modelled geoelectric field without affecting coherence. The simulation of the storms provides both intensity information as well as orientation at each individual site, adding $\approx 10-20~\%$ additional relative error/RMS compared to a standard non real-time method, which is still sufficient to monitor geoelectric activity (Fig. \ref{fig:Geo_Comp_2}). The choice of ideal correction curve varies slightly from site to site and storm to storm, depending on what frequencies the site is most sensitive to. Figure \ref{fig:Correct_Curve2} illustrates this difference in the correction curve obtained for each MT site, between the minimum and maximum extents. Due to this variability, the correction curve applied for the real-time model is not perfect, but does improve upon no correction (Fig. \ref{fig:Corr_factor}). However, almost no additional error is added, despite increasing the amplitude of the electric fields. Hence, the correction curve approach proved successful.
    
    \subsection{Galvanic Versus Non-Galvanic Distortion}\label{section:disc-galvanic}
    
    This study is the first attempt to apply a galvanic distortion correction, for space weather, across an entire region. Previous work by \citet{Murphy2021} indicated that disregarding local information can misrepresent the regional effects and can lead to an over/underestimation of the electric field in a local area. Here, we reach a similar conclusion, observing that galvanic corrections lead to more homogeneity in the direction of the electric field, which is likely to be more representative of the regional lithospheric structure. The distorted MT-TF also contains local information about the smaller near surface structures, which also generate electric fields and for certain sites can be up to an order of magnitude greater than the regional geoelectric field generated in the lithosphere. Individually, local near-surface effects are likely too weak to drive GIC on their own. As such, despite the relatively high density of MT sites used here (compared to other studies), one would still expect a model with a galvanic distortion correction to produce more temporally coherent results. However, a combination of local inhomogeneities, related to a real geological feature (i.e., a fault), could have a relevant role on induced GIC for a particular power-line or pipeline, given that near surface inhomogeneities can group together (in our case, we see these inhomogeneities are more common along the coastline and in some mountainous regions like the north-west, while the midlands remained very homogeneous).  For sparse MT networks, using galvanic corrections would most likely improve GIC modelling, however this is less clear if a high density of MT-TF were available and further study is needed to clarify this point.  While galvanic corrections provide a better regional picture, and are likely to be more representative and provide more coherent results in the long term, uncorrected galvanic distortion also needs to be considered to highlight potentially larger geoelectric fields in certain areas, than what is implied by the corrected galvanic distortion.
    
    \subsection{Limitations}\label{section:Limitations}
    \begin{enumerate}
        \item The plain wave approximation is likely to be more erroneous when local effects are present in geomagnetic storms. This could be solved by using a 3D model of the lithosphere in Ireland and its interaction with the ionosphere and magnetospheric currents. However, this approach would be more computationally expensive and would require modification. New research by \citet{Kruglyakov2022} presents a possible solution to this problem, by using a memory based method to reduce computation time.
    
        \item At present the magnetometer density is low. This leads to errors in the calculation of SECS at more isolated sites (such as site E, Fig. \ref{fig:Mag_MT_Sites}). This, in turn, can lead to under/over estimations in the amplitude of the modelled geoelectric fields. To mitigate this we have installed one new magnetometer in the east of Ireland in Dunsink Observatory, Co. Dublin, and we plan to install two more, one in the north-west and one in the south-east. In particular, a new site is required in the north-west, demonstrated by Figures \ref{fig:coh_snr1}, \ref{fig:coh_snr2}.

        \item The correction for galvanic distortion considered in this study does not account for the scalar factor. This means that although the impact of galvanic distortion is largely reduced, some distortion may still be present, affecting the amplitude of the electric field. 
    \end{enumerate}

\section{Conclusions}\label{section:conclusions}
    
    \begin{enumerate}
        \item{ An operational/automated geoelectric field monitoring approach has been successfully implemented to accurately estimate amplitude, orientation and uncertainties of near real time (1-minute) geoelectric fields in Ireland.}
        \item The nowcast model adds a greater uncertainty than standard modelling, with a coherence loss of $\approx~5~\%$ and an additional $5-10~\%$ in the uncertainties of the total amplitude.
        \item {Optimising the padded magnetic time series and applying a correction factor were crucial for real-time modelling with MT-TF, in particular the correction factor improved upon the previous state of the art for near real-time modelling of geoelectric fields.}
        \item {The effect of galvanic corrections on electric fields across the entire network was investigated. The correction led to lower amplitudes in electric field strength and more homogeneity in electric field direction. Using galvanic correction for MT-TF would likely lead to improved temporal coherence in GIC modelling.}
        \item {The real-time model can be used as a proxy to highlight regions or power-lines that are more likely to be affected by GIC in near real time, which is more relevant than simply looking at magnetic variations as a proxy.}
        \item {Modelled real-time electric field data is stored at \url{www.magie.ie}, where it is made freely available to the public.}
    \end{enumerate}

\section*{Data Availability and Supplementary Materials}
\begin{enumerate}
    \item{Magnetometer data used in this project is publicly available and found at \url{https://data.magie.ie}. }
    \item{The MT-TF from the SWEMDI database are currently available upon request to Geological Survey Ireland (research@gsi.ie) and will be made open access in the future. }
    \item {Live plots of nowcast geoelectric fields are available at \url{https://www.magie.ie/geoelectrics/}.}
    \item{The scripts for the nowcast model are located at \url{https://github.com/TCDSolar/NowcastGeoelectrics/}.
    \item{Example animations of the nowcast model are available at ...}
    }
    \begin{itemize}
        \item The nowcast model -- {nowcast\_model.mp4}
        \item The standard deviation of the nowcast model -- {nowcast\_model\_SD.mp4}
        \item The nowcast model with galvanic distotion -- {nowcast\_model\_galvanic\_dist.mp4}
    \end{itemize}
\end{enumerate}

\begin{acknowledgements}
    We would like to thank the Irish Research Council, Geological Survey Ireland and Dublin Institute for Advanced Studies for supporting this work as part of an IRC Enterprise Partnership Scheme. We also acknowledge Armagh Observatory and Trinity College Dublin for hosting magnetometers which contributed to this work, to Met Éireann for the Valentia Observatory magnetic measurements and INTERMAGNET for provided access to the other magnetometers used in this study. 
\end{acknowledgements}

\bibliography{jswsc}

\section*{Appendix}\label{sec:appendix}
    \subsection{Performance Metrics}\label{section:perf-metrics}
    The performance of the both the nowcast model was compared to the standard models results/measurements using relative error, root-mean-squared (RMS) error, signal-to-noise ratio (SNR) and coherence, for multiple time series. 
    
    A performance score or parameter P (first used by \citet{Torta2015}) was used to optimise the best correction curve for single time series. A performance score worked best when optimising a real-time correction factor, when compared to optimisation using solely RMS, SNR or coherence.
    
    \begin{equation}
        Rel = \frac{1}{N}\sum_{i=1}^{N}\frac{E_{data}-E_{mod}}{E_{data}}
        \label{eq:rel-error}
    \end{equation}
    
    \begin{equation}
        RMS = \sqrt{\frac{1}{N}\sum_{i=1}^{N}(E_{data}-E_{mod})^2}
        \label{eq:RMS}
    \end{equation}
    
    \begin{equation}
        SNR = 10 log_{10}\frac{\sum_{i=1}^{N}E_{data}^2}{\sum_{i=1}^{N}(E_{data}-E_{mod})^2}
        \label{eq:SNR}
    \end{equation}
    
    \begin{equation}
        Coh = \frac{\sum_{i=1}^N(E_{data}-\bar{E}_{data})(E_{mod}-\bar{E}_{mod})}{\sqrt{{\sum_{i=1}^N}(E_{data}-\bar{E}_{data})^2(E_{mod}-\bar{E}_{mod})^2}}
        \label{eq:Coh}
    \end{equation}
    
    \begin{equation}
        P=1-\frac{RMS}{\sigma_0}
        \label{Eq:perf}
    \end{equation}
    \mbox{}\\
    where $N$ is the length of the time series and $\sigma_0$ is the standard deviation

    The mean of each metric was also calculated across multiple time series, when comparing performance of the model at separate delays from real time. For example the mean of the RMS can be given by 

    \begin{equation}
        \overline{RMS}=\frac{\sum_{i=1}^{n}{RMS}}{n}
        \label{eq:meanRMS}
    \end{equation}

    where $n$ is the number of time series used.

    \subsection{Padding Types}

    \subsection{Performance of SECS with Kp Index}

\begin{figure}[H]
            \centering
        \includegraphics[scale=0.3]{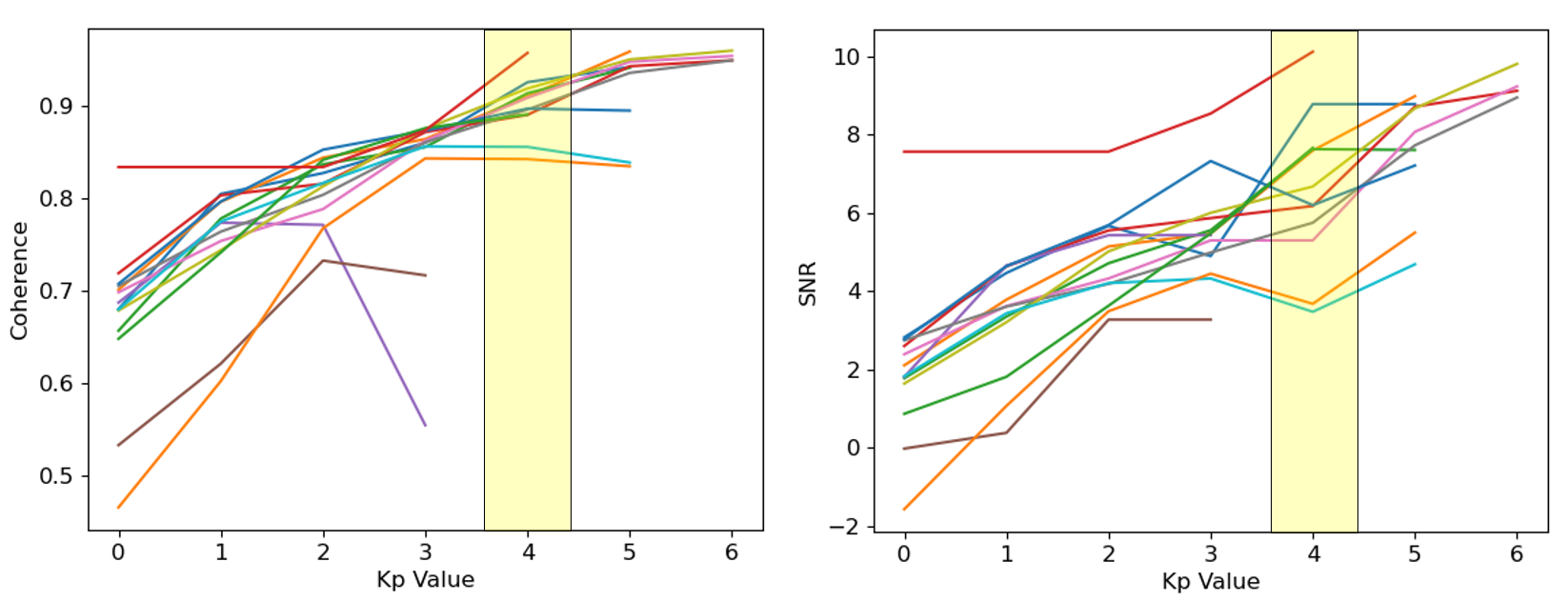}
        \caption{The mean of coherence (left) and SNR (right) comparing measured magnetic field data at newer SWEMDI MT sites with SECS interpolated data (using 3 sites). The SNR and coherence of this comparison is plotted against kp value, highlighting the increase in these metrics with greater magnetic activity. The SNR and coherence of each 3 hour Kp Index bin was recorded, with the mean value plotted. }
        \label{fig:kp_cohsnr}
    \end{figure}

    SECS-derived magnetic field data were compared with measured magnetic data at MT sites across varying magnetic activity. Figure \ref{fig:kp_cohsnr} gives a summary of the accuracy of the SECS interpolation using coherence and SNR.

\end{document}